\documentclass[prc,twocolumn,showpacs,superscriptaddress]{revtex4-1}
\usepackage{dsfont}
\usepackage{graphicx,amsmath}
\usepackage{xcolor}
\def\rev#1{\textcolor{red}{#1}}
\def\be{\begin{eqnarray}}
\def\ee{\end{eqnarray}}
\def\bc{\begin{center}}
\def\ec{\end{center}}

\newcommand{\lsim}{\stackrel{\scriptstyle <}{\phantom{}_{\sim}}}
\newcommand{\gsim}{\stackrel{\scriptstyle >}{\phantom{}_{\sim}}}

\begin{document}
\title{Pion condensation at rotation in magnetic field,  electric and scalar potential wells
}
\author{D. N. Voskresensky}
   \affiliation{ Joint Institute for Nuclear Research,
		Joliot-Curie street 6,
		141980 Dubna, Russia}
 \affiliation{ National Research Nuclear
    University (MEPhI), 115409 Moscow, Russia}
\begin{abstract}
Instability of the  charged pion vacuum resulting in the pion condensation is studied within the $\lambda|\phi|^4$ model at the rapid rotation of the system in presence of  external fields. Case of the static
external scalar and electric square potential wells and the uniform magnetic field is studied in detail.  The  Meissner and Aharonov-Bohm effects and the London moment are taken into account. Analogies and differences with the behavior of metallic superconductors under the action of the rotation and the external magnetic field are analyzed.
\end{abstract}
\date{\today}
 \maketitle

\section{Introduction}

In condensed matter physics the behavior of the rotating superfluid $^4$He and cold atomic gases was extensively studied, cf. \cite{LP1981,PethickSmith,PitString,BaymChandler1983,BaymChandler1986,Fischer:2003zz} and references therein. With the low angular velocity,  $\Omega < \Omega_{c1}$, the  superfluid $^4$He and cold atomic gases, being placed inside the initially resting vessel, do not respond on the subsequent rotation of the vessel, since production of elementary excitations and  vortices is energetically not profitable in this case. With increasing rotation frequency $\Omega$, for $\Omega > \Omega_{c1}$, the system produces   filament vortices of the normal matter immersed in the superfluid matter. Then, for $\Omega > \Omega_{\rm lat}>\Omega_{c1}$ the vortices form the triangular lattice, which mimics the rigid-body rotation of the vessel. For $\Omega >\Omega_{\rm c2}> \Omega_{\rm lat}>\Omega_{c1}$ the classical condensate field is completely destroyed.

Resting metallic superconductors respond  on action of the external uniform constant magnetic field $H$ similarly to the response of neutral superfluids on rotation, cf. \cite{LP1981,Abrikosov1988}. A low  magnetic field $H$ applied to the superconductor is screened on the so-called penetration depth of the magnetic field $l_h$ (inversed effective photon mass) near the boundary by the superconducting current occurring in this surface layer (Meissner-Higgs effect).
Superconductors are subdivided on two classes, superconductors of the first and second kind, in dependence of the value of the Ginzburg-Landau parameter $\kappa=l_h/l_\phi$, where $l_\phi$ is the so-called coherence length,
the scale of the change of the condensate field. For $\kappa<1/\sqrt{2}$ one deals with the superconductor of the first kind and for $\kappa>1/\sqrt{2}$, with the superconductor of the second kind. In the latter case for $H>H_{c1}\simeq H_{c}\ln \kappa/(\sqrt{2}\kappa)$ the surface energy associated with the formation of normal filament vortices becomes negative, $H_{c}$ is the so-called thermodynamic magnetic field, at which the volume energies of the phase with the equilibrium value of the condensate field $\phi =\phi_0$ and the magnetic field $\vec{h}=\mbox{curl}\vec{A}=0$ and the phase with $\vec{h}=\vec{H}$ and $\phi =0$ are equal. Vortices form the triangular Abrikosov lattice. For $H>H_{c2}\simeq \sqrt{2}\kappa H_c>H_{c1}$ the condensate is completely destroyed. In case of pions, the pion-pion interaction constant $\lambda\gg e^2=1/137$, $\hbar=c=1$, and we deal with the superconductivity of the second kind, cf. \cite{Voskresensky:1980nk,V88,Migdal:1990vm}.  Penetration of the magnetic field in the cylindrical  superconductors has been also studied,   cf.  \cite{LP1981,Zharkov2001} and references therein.

 Superconductors may generate  the uniform magnetic field in the interior,  decreasing   on the scale of the penetration depth $l_h$ from the boundary, which allows for the corotation of the charged superfluid and the oppositely charged normal matter component in the laboratory frame even in absence of vortices, for $\Omega<\Omega_{c1}$, cf. \cite{London,Essen2005,Tajmar2008,Hirsch2013,Hirsch2018,Wang2020}.  This effect is called the London law or the London moment.  Thereby in case of superconducting materials the  state corresponding to the  initially not rotating  charged electron superfluid in the laboratory frame and the resting normal ion  liquid in the rotation frame is  unstable on a  large
 timescale. The resulting value of the London magnetic moment depends on the shape of the body, cf. \cite{Essen2005}.

 The superfluid current density in nonrelativistic quantum mechanics
is given by the ordinary expression
\be &\vec{j}_s=\frac{iq}{2m}(\Psi(\nabla +iq\vec{A})\Psi^*+c.c.) \nonumber\\&\simeq \frac{q\tilde{n}_0}{m}(\nabla \Phi -q\vec{A})=q\tilde{n}_0\vec{W}_s\,,\label{curMacro}
\ee
where $c.c.$ denotes complex conjugation, and we used presentation  $\Psi=\sqrt{\tilde{n}_0}e^{i\Phi}$ for the macroscopic wave function,
$\vec{A}$ is the vector potential of the  magnetic field,  $\tilde{n}_0$ is the  local condensate density, which for $T=0$ coincides with the density of the superfluid component $n_s$, $q$ and $m$  are
 the  charge and mass of the effective boson responsible for the superfluidity/superconductivity.
  In case of metallic superconductors $q=2e$, $e$ is the charge of electron and $m=2m_e$, $m_e$ is the mass of electron. The second equality (\ref{curMacro}) holds for $\kappa \gg 1$ that we will assume to be fulfilled. Then $n_0\simeq const$  at the typical distances $r\gg l_\phi$ from the boundary. It is experimentally proven that in presence of the normal component of the current (e.g., produced by the ion lattice) in the laboratory frame
\be\vec{j}_{\rm ion}=q_{\rm ion}n_{\rm ion}\vec{W}_{\rm ion}=-qn_s\vec{W}_{s}\,,\label{London1}
\ee
 $n_{\rm ion}=n_s$ provided the system is electroneutral,  the created magnetic field  forces the superfluid and normal liquids to co-rotate. The superfluid velocity at the distance  $r_3=\sqrt{x^2+y^2+z^2}$ from the rotation axis of the cylindric body is given
by $\vec{W}_s =[\vec{\Omega}\times \vec{r}_3]$.
From (\ref{curMacro}) for $\nabla\Phi=0$ we obtain $q\vec{A}_{\rm L}=-m\vec{W}_s$. Also, after applying operator $\mbox{curl}$ to  Eq. (\ref{curMacro})  one obtains for irrotational motion
\be \nabla\times \vec{W}_s =-q\vec{h}/m\,.\label{LondonEq}
\ee
  Using that $\nabla\times [\vec{r}_3\times \vec{\Omega}]=2\Omega\,$  we recover the value of the London uniform magnetic field generated in the interior  of the cylindrical rotating superconductor
 \be
 \vec{h}_{\rm L} = -(2m/q)\vec{\Omega}\,.\label{London}
 \ee
 It is remarkable that
$\vec{h}_{\rm L}$ is directed along $\vec{\Omega}$ for $q<0$ and in opposite direction for $q>0$, whereas the ordinary Meissner current depends on $q^2$.

Also from integration of the expression (\ref{curMacro})  over a closed contour  using  the Stokes theorem one has
\be\oint q\vec{A}_{\rm L}d\vec{l}=\oint q\vec{h}_{\rm L}d\vec{S}=2\pi\nu-2\Omega mS\,,
\ee
where $\oint q\vec{h}d\vec{S}$ is the magnetic flux and $S=\pi r^2$, and $\nu$ is the integer winding number characterizing the vortex. In case of uniform magnetic field and irrotational motion ($\nu =0$) we again recover Eq. (\ref{London}).


Cooper pairing of nucleons is responsible for the decrease of the nuclei moments of inertia, cf. \cite{MigTFFS}, and  glitches  of pulsars, cf. \cite{ST83}. In the neutron star interiors neutrons are supposed to be paired in the $1S_0$ state at the baryon densities $n \leq (0.7-0.8) n_0$, where $n_0$ is the nuclear saturation density,  and in  $3P_2$ state ($L=S=1$ and $J=2$)  at a higher baryon density. Protons are paired in the $1S_0$ state. Response of these superfluids on the rotation affects properties of neutron stars,  cf. ~\cite{ST83,Sauls1989,Migdal:1990vm,Carter1999,Sedr2019}.

Pseudoscalar pions and scalar sigma mesons have the lowest masses among mesons. The pion condensates can exist in the neutron star interiors \cite{Migdal1978,Migdal:1990vm} and feasibly can  be produced at some conditions in  heavy-ion collisions, cf. \cite{Voskresensky1994,Pirner:1994tt,Voskresensky:2022fzk}. The charged pion condensate  behaves as a superconductor \cite{Voskresensky:1980nk,Migdal:1990vm}, the $\sigma-\pi^0$ chiral wave condensate may form a superfluid \cite{Voskresensky:2023znr,Voskresensky:2024ivv}.
Due to a strong p-wave pion-nucleon attraction the pion condensation occurs probably in the $p$-wave state  \cite{Migdal1978,Migdal:1990vm}, whereas the $s$-wave condensation is also possible in some models, cf.\cite{Voskresensky:2022gts}. Specifics of the superconductivity of the $p$-wave pion condensate was studied in \cite{Voskresensky:1980nk}.

Magnetic fields in ordinary pulsars, like  the Crab pulsar, reach values  $\lsim 10^{13}$ G at their surfaces.
At the surface of magnetars magnetic fields are higher, $\gsim 10^{15}$G. In the interior the magnetic field might be even stronger (up to $\sim 10^{18}$G) depending on the assumed (still badly known) mechanism of the  formation of the magnetic field~\cite{Ghosh}.
Also, magnetic fields $\gsim 10^{17}$G, $h\sim h_{\rm VA}=Z|e|/R^2\sim H_\pi (Ze^6)^{1/3}$ (where typical time of the collision process was estimated as $\Delta t\sim R/c$)  may exist in noncentral heavy ion collisions  at collision energies $\sim $ GeV/A, as it was  evaluated  in Ref. ~\cite{Voskresensky:1980nk}, $Z$ is the charge of the fireball, $H_\pi =m_\pi^2/|e|\simeq 3.5\cdot 10^{18}$G. For higher collision energies such  estimated value is in average enhanced by the dependence on the $\gamma =1/\sqrt{{1-v^2}}$ factor, cf. Eq. (1) in  ~\cite{IST}, reaching values  $h_{\rm SIT}\sim 10^{19}$G at $\sqrt{s_{NN}}=200$, that effectively corresponds to $h_{\rm SIT}\sim h_{\rm VA}\gamma$ ($\Delta t\sim R_\gamma=R/\gamma$
due to Lorentz contraction,
 cf. Eq. (38.8) in \cite{LL2} for the field in perpendicular direction to the motion of colliding nuclei). These  estimates do not include a suppression factor associated with the fact that rapidly varying fields can be described in the classical approximation only provided $E^2\sim H^2\gg \int^\omega k^3dk\sim \omega^4/4$, cf. \cite{LL4}, where $\omega\sim 1/R_\gamma$
 is the typical frequency of the field. Thus classical approximation holds provided $H\gg H_\pi \gamma^2 |e|/Z^{2/3}$. Comparing this value and $h_{\rm SIT}$ we see that classical approximation hardly holds  for $\gamma\gsim Z|e|$. Thus  for ultrarelativistic energies, for $\gamma\gsim Z|e|\gg 1$, we may estimate the  maximum value $h$, as  $h_{max}\sim h_{\rm VA} Z|e|$. Note that estimate of maximum value of the magnetic field extracted from the analysis of the ALICE Collaboration data  for 5.02 TeV Pb-Pb collisions in \cite{ALICE} coincides with simple estimate $h_{\rm VA}$,  whereas \cite{Vovchenko} estimated  $eh_{max}\simeq 6.3 m_\pi^2$.

There exist  fast-rotating neutron stars with the angular rotation  frequencies   $\lsim 10^4$ Hz, see Ref.~\cite{Ghosh}. When such a neutron star will undergo collapse, $\Omega$ will be increased due to conservation of the angular momentum and we estimate $\Omega< 1/r_G=1/(2MG)\sim 10^5$Hz for $M\sim M_{\odot}$. For primordial black holes with mass $M<10^{14}$g, which would not have survived to the present time due to the Hawking radiation,
this estimate yields $\Omega\lsim 10^{24}$Hz.

Rotational frequencies in nuclei usually do not exceed $\Omega\sim 3\cdot 10^{21}$Hz, cf. \cite{AfanasievNucl}.
Estimates yield angular momenta $L\sim \sqrt{s}Ab/2\lsim 10^6\hbar$ in peripheral heavy-ion collisions of Au $+$ Au  at $\sqrt{s} = 200$ GeV, for the impact parameter $b = 10$ fm, where $A$ is the nucleon number of the ion  \cite{Chen2015}. The global
polarization of $\Lambda (1116)$  hyperon observed  by the STAR Collaboration in noncentral Au-Au collisions  \cite{Adamczyk2017} indicated existence of a vorticity with  rotation frequency  $\Omega\simeq (9\pm 1) \cdot 10^{21}$ Hz $\simeq 0.05m_\pi$, $m_\pi\simeq 140$\,MeV is the pion mass. Employing (\ref{London}) for  pions we estimate  ${h}_{\rm L} \simeq 3\cdot 10^{-5} \Omega$, with ${h}_{\rm L}$ measured in Gauss and $\Omega$ in rad$/$s, i.e. we would have $h\sim {h}_{\rm L} \simeq 3\cdot 10^{17}$G for $\Omega\simeq  10^{22}$ Hz, if we dealt with the charged pion Bose-Einstein condensate.

At SPS, RHIC and LHC heavy-ion collision energies   at midrapidity a  baryon-poor
medium is formed  \cite{Afanasiev,Alt,Nayak,Abelev,Adamczyk} with pion number exceeding the baryon/antibaryon number more than by the order of magnitude.
It is commonly assumed that hadrons  are produced at the hadronization temperature
$T=T_{\rm had}$  after cooling of an expanding quark-gluon fireball, cf. \cite{Stachel}.
At the temperature of the chemical freeze-out, $T_{\rm chem}\simeq T_{\rm had}$,  inelastic processes are assumed to be ceased and  up to the thermal freeze-out predominantly elastic processes occur. Thus at the stage $T_{\rm chem}>T(t)>T_{\rm th}$ one may speak about a dynamically fixed pion number, i.e. approximately not changing at this time-stage. If the state formed at the chemical freeze-out was   overpopulated, during the cooling there may appear the Bose-Einstein  condensate  of pions   characterized by the dynamically fixed pion number, as it was suggested  in  \cite{Voskresensky1994}. On the time interval, at which the particle number remains almost constant, the Bose-Einstein  condensate  of pions  behaves similar to a superfluid.
 A nonequilibrium Bose–Einstein condensation  has been observed and studied  in case of the so-called exiton-polariton bosonic quasiparticles that exist inside semiconductor microcavities  \cite{Exiton2010,Byrnes2014}.
The ALICE Collaboration observed a significant suppression of three and four pion
Bose–Einstein correlations in Pb-Pb collisions at
$\sqrt{s_{NN}} =2.76$ TeV at the LHC \cite{Abelev2014,Adam2016}. This circumstance can be interpreted as there is a considerable degree of coherent pion
emission in relativistic heavy-ion collisions \cite{Akkelin2002,Wong2007}. Analysis \cite{Begun2015} indicated that about $5\%$ of pions could stem from the Bose-Einstein condensate. A discussion of a possibility of the  Bose-Einstein condensation  in heavy-ion collisions at LHC energies can be found in the reviews \cite{Shuryak2017,Voskresensky:2022fzk}. Behavior of the charged pion and $\sigma-\pi^0$ Bose-Einstein condensates with the dynamically conserved particle number under the rotation at zero temperature was recently studied in \cite{Voskresensky:2023znr,Voskresensky:2024ivv}.

The problem of instability of  the pion vacuum in the rotation frame under action of external fields was raised  in  \cite{Zahed,Guo,Voskresensky:2023znr,Voskresensky:2024ivv}.  Works \cite{Voskresensky:2023znr,Voskresensky:2024ivv} considered possibilities of production of
the charged pion and neutral $\sigma-\pi^0$ condensates in ultrarelativistically rotated systems in presence or absence of the  rectangular potential well for the zero-component of the vector field, e.g. electric field in case of charged pions.   An analogy  was demonstrated  with the description of the nonrelativistic Bose-Einstein condensates  with a fixed particle number under the rotation. The main difference is that in the former case (case of vacuum) the chemical potential $\mu$ either equals zero or $-m_\pi$ and in the latter case (case of nonrelativistic Bose gas) $\mu\simeq m_\pi$. In case of the vacuum, e.g., in presence of a rather deep potential well for $\pi^-$ and the rotation, the ground state  $\pi^-$ level  may cross the boundary of the lover continuum, then $\pi^-$s occupy this level and $\pi^+$s with energy $+m$ go off to infinity.  In presence of a positively charged matter the $\pi^-$s can   be produced also    in reactions $X\to X+\pi^-+e^+ +\nu$, when the $\pi^-$s level reaches energy $-m_e\simeq 0$, where $m_e$ is the electron mass. In case when in the scalar and electric potential wells the lowest energy level reaches zero (or $-m$) even in absence of the rotation, the condensate arises from the vacuum both in the laboratory and the rotation frames even for $\nu=0$ at a weak rotation. It is then described  similarly to the order parameter  in ordinary superconductors. If $\mu$ can reach zero (or $-m$) only in presence of the rotation, the condensation does not occur in the  vacuum described in the laboratory frame, whereas the vortex charged pion field with the winding number $\nu\gg 1$ can be formed in the rotation frame.
References \cite{Zahed,Guo} suggested occurrence of the charged pion condensation in the rotation frame at the relativistic rotation in the external uniform magnetic field, such that $N=|eH|R^2/2\gg 1$, where $N$ is the degeneracy of the level and $R$ is an artificially chosen transverse radius satisfying   the causality condition $R<1/\Omega$. In the  case, which the authors considered, the instability may appear for $\Omega N/2 >m$ (for $|eH|\ll m^2$). Then  the particle  energy is $-\Omega N +m=-m$ and the antiparticle is produced with the energy $m$.

The important point  is that  the uniform external magnetic field  should not penetrate the region of the charged pion condensate owing to the Meissner effect provided $H<H_{c1}$. For $H_{c2}>H>H_{c1}$  the Abrikosov lattice can be formed, as it was explained above. Thereby, only the field $H\simeq H_{c2}$ can be considered as quasiuniform. Influence of the electric field was not considered  in \cite{Zahed,Guo}. The question about   possibility of appearance of the London moment also was not studied.

Subtlety  of the problem is that the  production of the field and particles depends on the noninertial reference frame, in which the phenomenon is described. In a sense this phenomenon is similar to the Unruh effect resulting in production of soft photons in the unevenly moving frame.

In the given paper we return to the problem of the stability of the charged pion vacuum  at the rotation of the system in presence and absence of the uniform external magnetic field. Also, presence of the square scalar and electric potential wells  will be permitted. Analogies and differences with the problem of the rotation of the charged pion Bose-Einstein condensate with dynamically fixed particle number and the rotating superconductors will be analyzed.

The paper is organized as follows. In Sect. \ref{chir} we introduce description of  the  charged boson (pion) field in the rotation frame under action of static  scalar,  electric and magnetic fields.  Problem of the charged pion vacuum instability in the rotation frame  at ignorance of the Meissner effect and London moment will be studied  in Sect. \ref{sect-ignor}. The Aharonov-Bohm effect and consequences, which follow from the absence of the current in the rotation frame, will be considered in Sect. \ref{Aharonov-sect}. In Sect. \ref{sect-London} we will study  peculiarities of the  Meissner effect and the London law in relativistic region.  Appendix \ref{sect-rotSKG} discusses subtlety of the transition from the  Klein-Gordon-Fock equation in the rotation frame to the Schr\"odinger equation in the nonrelativistic limit. Some details of calculations are deferred to Appendices \ref{Assumptions}, \ref{AppendixA} and \ref{AppendixB}.

\section{Complex scalar field placed in external fields in rotation frame
}\label{chir}
\subsection{General considerations}
Let us  study behavior of a complex scalar field  in a rigidly rotating cylindrical system at the constant rotation frequency $\vec{\Omega}\parallel z$.
In cylindrical system of coordinates $(r,\theta, z)$ one has $\nabla =(\partial_r, \partial_\theta/r, \partial_z)$,  $r=\sqrt{x^2+y^2}$. The coordinate transformation between the laboratory $(t',\vec{r}{\,'})$ frame and the rotating  $(t,\vec{r})$ frame is as follows:  $t'=t$, $x'=x\cos (\Omega t)-y\sin (\Omega t)$, $y'=x\sin (\Omega t) +y\cos (\Omega t)$, $z'=z$. Employing it and that $(ds')^2=\delta_{\mu\nu}dx^{\prime\mu} dx^{\prime\nu}=(ds)^2$, where  $\delta_{\mu\nu}=\mbox{diag}(1,-1,-1,-1)$, one   recovers
 expression for the interval in the general rotating frame,
  \be (ds)^2=(1-\Omega^2 r^2)(dt)^2 +2\Omega y dx dt  -2\Omega x dy dt -(dr_3)^2\,,\nonumber
  \ee
 $r_3= \sqrt{r^2 +z^2}$.   Thereby, a rigidly rotating system must be finite, otherwise  the causality condition, $\Omega r<1$, is not fulfilled. The tetrad is determined as \cite{Chen2015}:
 \begin{eqnarray}
 e_{\hat{\alpha}}^{\,\,\,\mu}=\delta_{\hat{\alpha}}^{\,\,\,\mu}-\delta_{\hat{\alpha}}^{\,\,\,0} \delta_i^{\,\,\mu} W^i\,,
 \label{tetrades}
 \end{eqnarray}
 $\vec{W}=(0, W^1,W^2,W^3)=[\vec{\Omega}\times \vec{r}_3]$, i.e.
 $e_{\hat{0}}^t=e_{\hat{1}}^x=e_{\hat{2}}^y=e_{\hat{3}}^z=1$, $e^{\,\,x}_{\hat{0}}=y\Omega$, $e^{\,\,y}_{\hat{0}}=-x\Omega$,
  other elements are zero.  Latin index $i=1,2,3$, greek tetrad and Lorentz indices $\hat{\alpha},\mu =0,1,2,3$; $e_{\hat{\alpha}} =e_{\hat{\alpha}}^{\,\,\mu}\partial_\mu$, $e_{\hat{0}}=\partial_t +y\Omega\partial_x -x\Omega\partial_y$, $e_{\hat{i}} =\partial_i$.
Thus in the
local flat frame we should  perform  the replacement
\be\partial_t\to \partial_t +y\Omega \partial_x -x\Omega \partial_y=\partial_t -i\Omega\hat{l}_z=
\partial_t -\Omega \partial_\theta\,,\label{shiftrotelectric}\ee
$\Omega\hat{l}_z=\vec{\Omega}[\vec{r}_3\times \hat{\vec{p}}]$, $\hat{\vec{p}}=\nabla/i$.

 The angular momentum associated with the charged pion field $\phi$  is
\begin{eqnarray}
&{\vec{L}}_{\pi}=\int d^3 X [ {\vec{r}}_3\times{\vec{P}}_\pi]\,\,,
\nonumber\\
&{P}^i_\pi=T^{0i}_\pi=-\frac{\partial{\cal{L}}}{\partial\partial_t \phi}\nabla_i \phi -\frac{\partial{\cal{L}}}{\partial\partial_t{\phi}^*}\nabla_i \phi^*,\,\,\label{pphi}
\end{eqnarray}
$T^{0i}_\pi$ is the $0i$ component of the energy-momentum tensor for the pion subsystem.

In presence of the external scalar field $U_{\rm sc}$ and the gauge field $A_\mu$ in the rotation reference frame, the part of the Lagrangian density corresponding to  the self-interacting complex scalar  field $\phi$ (to be specific we will further speak about the charged pion field) measured in the rotation frame renders \cite{Chen2015,Zahed,Guo}:
\begin{eqnarray}
&{\cal{L}}_\pi={|(D_t+y\Omega D_x -x\Omega D_y)\phi|^2} -{|D_i \phi|^2}\nonumber\\
&-{m^{*2} |\phi|^2}-\frac{\lambda |\phi|^4}{2}\,,\label{LVrot}
 \end{eqnarray}
where $D_\mu =\partial_\mu +ieA^{\rm fl}_\mu$, $eA_{\hat{\mu}}^{\rm fl}=e_{\hat{\mu}}^{\,\,\,{\nu}}eA_{{\nu}}$,  $e$ is the charge of the electron,
\be m^{*2}=m^2+U_{\rm sc}\,,
\ee
$m$ is the mass of the pion, $U_{\rm sc}$ is the external scalar potential, $m^*$ has the sense of  the effective mass of the pion, and in case of the cylindric symmetry under consideration we take $A_\nu (x,y)$ in the form  $A_\nu= (A_0, -A_x, -A_y,0)$, $A^{\rm fl}_\mu (x,y)$ is the vector potential in the local flat frame expressed in terms of the variables of the rotation frame.  Then we have
 \be eA_\mu^{\rm fl}=(eA_0-eA_x\Omega y+eA_y x\Omega, -A_x, -A_y, 0)\,,
\ee
and thereby
\be
D_t+y\Omega D_x -x\Omega D_y=\partial_t +ieA_0 +y\Omega \partial_x -x\Omega \partial_y\,.\label{Genshift}
\ee
In our exploratory study we will employ the simplest $-\lambda |\phi|^4/2$ charged pion self-interaction term, with  constant   $\lambda \sim 1$, which allows for analytical consideration. Within  the chiral $\sigma$ model, or in the model employing the Weinberg interaction,  or  in the low density limit within the general chiral perturbation theory, the effective pion-pion interaction becomes function of the pion frequency and momentum and the density of the medium, e.g. cf. \cite{Migdal:1990vm,BraunerYamamoto,Yamamoto,Evans:2022hwr}. The $-\lambda |\phi|^4/2$ term can be found with the help of the expansion of the interaction part of the Hamiltonian in the parameter $|\phi|/f_\pi\ll 1$, where $f_\pi\simeq 92$ MeV is the pion decay constant. Due to the mentioned pion energy-momentum and density dependence, $\lambda_{\rm ef}$ may even change the sign  in some region of the parameters. In this case the expansion should be continued at least up to the $|\phi|^6$ term.

The total Lagrange density is as follows, ${\cal{L}}={\cal{L}}_\pi+{\cal{L}}_{\rm el}$,
\begin{eqnarray}
{\cal{L}}_{\rm el}=-\frac{F_{\rm fl}^{0i}F_{0i}^{\rm fl}}{8\pi}-\frac{F^{ij}_{\rm fl}F_{ij}^{\rm fl}}{8\pi}-j^\mu_{\rm ex,fl} A_\mu^{\rm fl}\,,\label{L0i}
\end{eqnarray}
  ${j}_{\rm ex, fl}^\mu =(-e n_{\rm ex}^{{\rm fl}},\, \vec{j}_{\rm ex}^{\,\,\rm fl})$ is the external current in the local flat frame produced by  particles heaving the charge $-e$, $j^\mu_{\rm ex} =(-en_{\rm ex}, \vec{j}_{\rm ex})$, $j^\mu_{\rm ex, fl} A_\mu^{\rm fl}=j^\mu_{\rm ex} A_\mu$. In the rotation frame $F^{\mu\nu}=\partial^\mu A^\nu - \partial^\nu A^\mu$, and in the local flat
  frame, cf. \cite{Xu},
\begin{eqnarray}
&F_{0i}^{\rm fl}=\partial_0 A_i -\partial_i A_0 -W_j \partial_j A_i -\epsilon_{ijk}\Omega_k A_j\,,\label{F0i}\\
&F_{ij}^{\rm fl}=\partial_i A_j- \partial_j A_i\,.\nonumber
\end{eqnarray}
Here  $\epsilon_{ijk}$ is the Levi-Civita tensor.
The Lagrangian density is split into the rotation independent part, ${\cal{L}}_{\rm el}(\Omega =0)$, and the rotation dependent one, cf. \cite{Xu},
\be \delta{\cal{L}}_{\rm el}(\Omega)=-\frac{W_j F_{ji}F_{0i}}{4\pi}+
\frac{1}{8\pi}\left(W_j\partial_jA_i +\epsilon_{ijm}\Omega_m A_j\right)^2.\label{deltaLi}
\ee

Let us focus  on  the case of the static  fields. We seek solution of the equation of motion in the form of the individual vortex:
\be \phi =\phi_{0}\chi (r)e^{i\xi (\theta)-i\mu t+ip_z z}\,,\label{Phifieldformpi}
\ee
where $\phi_{0}$,   $\mu$ and $p_z$ are  real constants,  $\chi(r)$ and $\xi (\theta)$ are  the real functions. Below we will disregard a trivial dependence  $\phi\propto e^{ip_z z}$ on $z$, since  we will be interested in description of the minimal energy configurations corresponding to $p_z=0$.

Using Eq.  (\ref{Phifieldformpi}) we  present the pion Lagrangian density   as
\begin{eqnarray}
{\cal{L}}_\pi={|\widehat{\widetilde{\mu}}\phi|^2 }-{|(\partial_i+ieA_i)\phi|^2}-{m^{*\,2} |\phi|^2}-\frac{\lambda |\phi|^4}{2}\,,
\label{LphiOmpiV} \end{eqnarray}
where
\begin{eqnarray}
\widehat{\widetilde{\mu}}=\mu  +iy\Omega\partial_x-ix\Omega\partial_y-V(r)\,,\label{murep}
\end{eqnarray}
$eA_0=V(r)$ is produced by the static external positive charge density $n_{\rm ex}({{r}})=n_{\rm ex}^{\rm fl}({{r}})$ and may be also by a redistribution of the charge owing to the rotation.  In cases when the electric and magnetic fields can be treated as external fields, only the term ${\cal{L}}_\pi$ remains, describing the charged pion superconducting subsystem. Otherwise,  the electromagnetic potential should be found self-consistently. In this respect we will further bear in mind  two possibilities, rotation of the empty electrically neutral vessel heaving internal radius $R<1/\Omega$ and rotation of a piece of the charged nuclear matter of the cylindric form.  In the latter case, simplifying the problem we will assume that the source $n_{\rm ex} (r)\neq 0$  is produced by very heavy particles, e.g., if we deal with a rotating piece of the  bounded nuclear matter and pions, $n_{\rm ex}=n_p$ is the density of protons, being 7 times heavier than pions. In the latter case we  put $n_p= n_p^{\rm rot}\simeq n_p^{\rm fl}$ and  we take $n_p=n_p^0\theta (R-r)$, and   $n_p^0=const>0$,  $\theta(x)$ is the step function, $R$ is the transversal size of the system. Due to the causality condition we employ  $R<1/\Omega$.

Circulation of the $\xi(\theta)$-field yields $\oint d\vec{l}\nabla \xi =2\pi \nu\,,$
at the integer values of the winding number $\nu =0,\pm 1,...$, thereby
\be \nabla \xi =\nu /r\,, \quad\xi =\nu\theta\,.\label{circul}\ee
 Positive $\nu$ characterize the vortices and negative ones, the antivortices. Employing Eq. (\ref{Phifieldformpi}) we obtain  $L_z^\pi =\int d^3x n_\pi \nu$, where $n_{\pi}=n_{\pi}^{\rm fl}$ is the charged pion field/particle density. Note here that namely the canonical angular momentum (\ref{pphi}) is conserved quantity, cf. \cite{Barnet}. Sense of the difference of the canonical and kinetic momenta is  discussed in \cite{Barnet,Chernodub2017}.

Equation of motion  for the condensate field is as follows, cf. \cite{Zahed,Guo,Voskresensky:2023znr,Voskresensky:2024ivv},
\begin{eqnarray}
&[\widetilde{\mu}^2
+(\nabla -ie\vec{A})^2  -m^{*\,2}]\phi\nonumber\\
 &-\lambda |\phi|^2\phi =0\,,\quad \widetilde{\mu}=\mu  +\Omega\nu-V(r)\,,\label{pipitot}
 \end{eqnarray}
 where we employed the replacement $\phi\to \phi e^{i\nu\theta-i\mu t}$.
Using Eqs. (\ref{L0i}), (\ref{F0i}), (\ref{deltaLi}) and (\ref{LphiOmpiV}) we recover  equation  for the electric field
\begin{eqnarray}
\Delta V\simeq 4\pi e^2 (n_{\rm ex}-n_\pi),\,\, n_{\pi}=-\frac{\partial{\cal{L}_\pi}}{\partial V}
=2\widetilde{\mu}|\phi|^2\,,\label{densmod1rot}
 \end{eqnarray}
 and the equation  for the vector potential,
\begin{eqnarray}\Delta \vec{A}=-4\pi (\vec{j}_{\rm ex}+\delta \vec{j}+\vec{j}_\pi)\,,\quad \mbox{div}\vec{A}=0\,,\label{Meis}
 \end{eqnarray}
where
 \begin{eqnarray}\vec{j}_\pi=\frac{\partial{\cal{L}}_\pi}{\partial\vec{A}}=ie\phi (\nabla+ie\vec{A})\phi^*+c.c.\,,\label{current}
  \end{eqnarray}
  and for $\vec{\Omega} \parallel z$, $A_3=0$, and for the static fields $A_i$ performing the variation of (\ref{deltaLi}) one obtains in the gauge $\partial_i A_i=0$, cf. \cite{Xu},
 \begin{eqnarray}
-4\pi\delta {{j}}_i=W_j\partial_j(W_n\partial_n A_i +2\epsilon_{inm}\Omega_m A_n)-\Omega^2 A_i\,.
   \end{eqnarray}
  After averaging over the $x,y$ directions we have   $4\pi\overline{\delta {j}}_\theta\simeq -\frac{1}{2}\Omega^2 r^2\Delta {A}_\theta(r)+\Omega^2{A}_\theta(r)$. The first term yields the rotational contribution to the magnetic susceptibility $\chi_{\rm mag}$, whereas the second one results in   an anti-screening contribution to the current. In nonrelativistic limit the $\Omega^2$ terms can be dropped and we can put $\delta \vec{j}=0$.
  Such a terms may appear in case of ferromagnetic superconductors described by the vector complex fields, cf. \cite{Voskresensky:2019zcp}.
   At the rapid rotation   of the system we  may  neglect the anti-screening contribution to $\delta \vec{j}$ compared to the term (\ref{current}) at the condition
  \be 8\pi e^2|\phi|^2\gg 1/R^2\,.\ee
  Further, simplifying consideration  we will also assume that  $\chi_{\rm mag}=1$  and  thereby we put $\delta \vec{j}=0$.

Let us notice that for $\lambda=0$  the Klein-Gordon-Fock equation  (\ref{pipitot}) differs only by the constant energy shift $\mu\to \mu+\Omega\nu$ from the Klein-Gordon-Fock equation describing stationary states of the charged pion with quantum number $\nu$ in the laboratory frame, e.g. cf. \cite{V88,Migdal:1990vm}.

The  energy density  is given by expression
\begin{eqnarray}
&E_\pi=\mu n_\pi  -{\cal{L}}_\pi\,.\label{EnerPhipi}
\end{eqnarray}
 Thus employing (\ref{LphiOmpiV}) we find the pion contribution to the energy density:
\begin{eqnarray}
&E_\pi=\mu n_\pi+|\nabla \phi|^2 -({\widetilde{\mu}}^2 -m^{*2})|\phi|^2+\lambda|\phi|^4/2\nonumber\\
&-\vec{j}_\pi\vec{A}-e^2\vec{A}^{\,2}|\phi|^2
\,,\label{EAsquare}
\end{eqnarray}
cf. Eqs. (\ref{LphiOmpiV}), (\ref{murep}), (\ref{pipitot}), (\ref{EnerPhipi}). As it is seen from Eqs. (\ref{Meis}), (\ref{current}), the term $-\vec{j}_\pi\vec{A}\sim |\phi|^4$ in (\ref{EAsquare}) and $-e^2\vec{A}^{\,2}|\phi|^2\sim |\phi|^6$ for small $|\phi|^2$.

 In case of the uniform constant external magnetic field $\vec{H}^{\rm rot}=\vec{H}\parallel z$ and electric field $eA_0^{\rm rot}(r)=V(r)$ we have
\be eA_\mu^{\rm fl}=(V(r)-eH\Omega r^2/2, eHy/2,-eH x/2,0)\,,\label{Arot}
\ee
for $\vec{A}^{\rm \,rot}=\vec{A}=(-Hy/2, Hx/2, 0)$. In this case  one should perform the replacement  $-\vec{h}^2/(8\pi) +\vec{A}\,\vec{j}_{\rm ex}\to-H^2/(8\pi)$ in (\ref{L0i}).
Equation of motion for the condensate field renders, cf. \cite{Zahed,Guo,Voskresensky:2023znr,Voskresensky:2024ivv},
\begin{eqnarray}
&[\widetilde{\mu}^2+\hat{K} -m^{*\,2}]\chi
 -\lambda |\phi_0|^2\chi^3 =0\,,\label{pipi}\\ &\hat{K}=\Delta_r  -\nu^2/r^2+eH\nu -(eH)^2r^2/4\,,\nonumber
 \end{eqnarray}
 $\Delta_r=\partial_r^2 +{\partial_r}/{r}$.

One should notice that  solutions found in  Refs.  \cite{Zahed,Guo} are  valid for description of the pion vacuum in the  local flat frame  in presence of the external uniform magnetic field only provided  the uniform  magnetic field is the  approximate solution of Eq. (\ref{Meis}). As we shall argue below, it is so only for $H$ varying in the vicinity of the value $H_{c2}$, cf. Eq. (\ref{hc2}) below.
Otherwise one should take into account the Meissner effect.

Some subtleties associated with the description of the nonrelativistic rotation are discussed in Appendix  \ref{sect-rotSKG}.

\subsection{Boundary conditions}

References \cite{Zahed,Guo,Voskresensky:2023znr,Voskresensky:2024ivv} employed  the boundary conditions
\be  \phi (r=R)=0\,,\label{boundcond}
\ee
and
\be \phi (r\to 0)\to 0\,, \label{boundcond0}
\ee
to describe the pion vortex with $\nu\neq 0$ in the external uniform magnetic field described by the vector potential (\ref{Arot}) in the
local flat frame.

The following remark is in order.   When  one formally considers  the problem in the comoving frame, there appears a problem with the treatment of the distances $r>1/\Omega$. Studying the case  $N=|eH| R^2\gg 1$, Refs. \cite{Zahed,Guo} considered $R$ as an arbitrary chosen  value (fixed by the chosen value $N$) satisfying however the causality condition $R<1/\Omega$. However it looks rather strange, if the physically meaningful result depends on an arbitrary chosen value $R$.
 Thereby   Refs.  \cite{Voskresensky:2023znr,Voskresensky:2024ivv} suggested to treat the rotation frame, as the rigidly rotating vessel putting  $R$ to be equal to  the internal radius of the vessel, $R<1/\Omega$. In the problem  of the rotating piece of the nucleon matter (of the cylindric form),  $R$ was chosen  to be equal to the transversal radius of the cylinder. In the theory of superconductivity of metals treating $R$ as the size of the cylindric body one often  chooses  the   boundary condition
 \be[\vec{e}(-i\nabla-q\vec{A})\phi|]_{r=R} =0\,,\label{boundder}
 \ee
  or a more general (Robin) boundary
condition
\be[\vec{e}(-i\nabla-q\vec{A})\phi]_{r=R} =i\phi_{r=R}/l\,, \quad \vec{e}=\vec{r}/r\,,\label{boundderRob}
\ee
 $l$ is a constant  determined from some additional conditions, cf. \cite{LP1981}. For $l\to 0$ we recover the Derichlet condition (\ref{boundcond}). For $l\to \infty$ we get the condition (\ref{boundder}).

 The boundary conditions for the order parameter should be supplemented by the corresponding boundary  conditions for $A_\mu$. For example, in case of the uniform magnetic field one can assume that $A^\prime_\theta (r=R)=H$. For $r<R$ to find $A_\theta (r)$ one should then solve the Maxwell equation (\ref{Meis}), since the uniform magnetic field may not satisfy it for $\phi\neq 0$.

\section{Vacuum instability in rotating system  at ignorance of Meissner effect and London moment}\label{sect-ignor}
\subsection{Case $H\to 0$ in absence of self-interaction}\label{HzeroNonint}

As in \cite{Voskresensky:2023znr,Voskresensky:2024ivv}, let us first put $\lambda =0$ and $\vec{A}=0$ in Eq. (\ref{pipitot}) ignoring thereby possible redistribution of the   magnetic field inside the system determined by Eq.
(\ref{Meis}). In this case  the equation of motion  for the condensate field (\ref{pipitot}) in the
local flat frame gets the form
 \begin{eqnarray}
[\Delta_r  -\nu^2/r^2 +\widetilde{\mu}^2-m^{*\,2}]\chi=0\,. \label{pipi2}
 \end{eqnarray}
Note that similarly  one neglects the  $\vec{A}$ dependence in the Ginzburg-Landau equation for  the order parameter in case of a static nonrelativistic superconductor placed  in a rather weak external magnetic field, cf. \cite{LP1981}. Also, the same Eq. (\ref{pipi2}) holds in case of neutral nonrelativistic and relativistic superfluids (in the limit $\lambda\to 0$), cf. \cite{LP1981,Voskresensky:2023znr}.
 The term $\nu^2|\phi|^2/r^2$ appearing in the energy density is associated with the centrifugal force. The term $-\Omega^2\nu^2|\phi|^2$ in expression for $\widetilde{\mu}^2$ yields the relativistic correction associated with the centripetal force. The cross term $-2\Omega\nu (\mu-V)|\phi|^2 $ is related to the Coriolis force. The term $\Omega\nu$ in the Klein-Gordon-Fock equation can be treated as a constant shift of the electric potential $V$.


Let $V=-V_0=const$ for $r<R$ is the external electric potential and let us neglect the redistribution of the electric charge inside the system. For  $\lambda =0$ the solution of the equation of motion  for the condensate field (\ref{pipi2}) satisfying the boundary conditions (\ref{boundcond0}), (\ref{boundcond}) is given by  the  Bessel function
 \begin{eqnarray}
 \chi (r)=J_\nu (r/l_\phi)\,,\quad l_\phi=1/\sqrt{\bar{\mu}^2-{m}^{*2}}\,, \quad \nu>0\,, \label{Bessel}\end{eqnarray}
 $\bar{\mu}=\mu+\Omega\nu+V_0$.
 This solution holds both for  $m^{*2}<0$, and also for $m^{*2}>0$ at  ${\mu}^2>{m}^{*2}$.

 For  $m^{*2}<0$ the vacuum is unstable for creation of the charged pion field both  in the laboratory and rotation reference frames already for $V=0$ and $\Omega=0$,  at $\nu=0$. Stability is provided by the self-interaction, $\lambda>0$, cf. \cite{V88,Migdal:1990vm}.

 Employing the boundary condition $\chi (R)=0$ we obtain the dispersion relation \cite{Voskresensky:2023znr,Voskresensky:2024ivv}:
 \begin{eqnarray} &\mu =-\Omega\nu -V_0
 +\sqrt{{m}^{*2}+j_{n,\nu}^2/R^2}\,,
 \label{grlev}\end{eqnarray}
where $j_{n,\nu}$ is the root of the Bessel function. Note that a formally  similar expression was derived in \cite{Yamamoto} for description of the isospin asymmetric quark matter under rotation, electric field effects were not considered there.  For $x=r/l_\phi\gg \nu$ we have $J_\nu (x)\simeq \sqrt{2/(\pi x)}\cos (x-\pi\nu/2-\pi/4)$.
From here we find approximate asymptotic value $j^{\rm as}_{n,\nu}\simeq \pi(\nu +1/2+2n-1)/2$, where the integer number
$n\geq 1$ is the corresponding zero of the function $\cos (x-\pi\nu/2 -\pi/4)$, and thereby $j_{1,1}^{\rm as}\simeq 5\pi/4\simeq 3.927$ that only slightly differs from the exact solution $3.832$.
For $\nu\gg 1$ we have $j_{1,\nu}^{\rm as}\to \nu+1.85575\nu^{1/3}$, e.g.,
$j_{1,10^4}\simeq 10040$.
Integral $\int^x x'dx' J^2_\nu (x')$ diverges at $x\to \infty$ and at least by this reason the restriction by $r<R$  is necessary, and usage of a  boundary condition for some $r=R$ is required.

For ${m}^{*2}<-j_{1,1}^2/R^2$  we deal with the instability, since $\phi\propto e^{+\sqrt{|{m}^{*2}+j_{1,1}^2/R^2|}t}\to \infty$ at arbitrary $\Omega$ including $\Omega =0$. Stability is recovered, if one takes into account the term $\lambda >0$, cf. \cite{V88,Migdal:1990vm} and Sect. \ref{HzeroSelfint} below.

 Let us focus attention on the case $m^{*2}>0$. The solution was studied in
   \cite{Voskresensky:2023znr,Voskresensky:2024ivv}.
Setting $\mu =0$ we find the value of the critical angular velocity
 \be\Omega_c =\left(\sqrt{m^{*2}+j_{1,\nu}^2}-V_0\right)/\nu\label{critcentr}\,.\ee
For $1\leq \nu =c_1 {m}^*R\ll {m}^*R$, i.e. at $c_1\ll 1$,  ${m}^*R\gg 1$ we have $\Omega_c =\Omega (\mu=\epsilon_{1,\nu}=0, c_1\ll 1)\simeq ({m}^*-V_0)/\nu >0$ and the vortex solution holds  for $V_0>V_{0c}\simeq m^*(1-c_1)$. In a more interesting case  $\nu=c_1{m}^* R\gg {m}^* R\gg 1$ (at $c_1\gg 1$) from (\ref{grlev}) we have
\begin{eqnarray}
&\mu=\epsilon_{1,\nu}\simeq -V_0+(-\Omega R +1)\nu/R+R{m}^{*2}/(2\nu)\nonumber\\&+1.86\nu^{1/3}/R+...\label{largenulim}
\end{eqnarray}
Setting in (\ref{largenulim})  the  limiting value $\Omega^{\rm caus}=1/R$ we get
\begin{eqnarray}
\epsilon_{1,\nu}\to -V_0+\frac{m^*}{2c_1}+\frac{1.86c_1^{1/3}m^*}{(m^*R)^{2/3}}...
\end{eqnarray}
 Thus the level $\epsilon_{1,\nu}$ may reach zero for $\Omega R<1$ at
 \be
 V_0>V_{0c}={m}^*/(2c_1)(1+O[(c_1^2/({m}^*R))^{2/3}])\label{Vc1}\ee
 for $1\ll c_1\ll \sqrt{m^*R}$.
For $\sqrt{m^*R}\gg c_1\gg 1$  formation of the supervortex (with a large $\nu$) becomes energetically favorable at
\begin{eqnarray}
&\Omega>\Omega_c =\Omega (\epsilon_{1,\nu}=0,c_1\gg 1)\simeq \frac{1}{R}-\frac{V_0-{m}^*/(2c_1)}{c_1{m}^*R},\label{condVom}\\ &V_0>V_{0c}={m^*}/{(2c_1)}\,. \nonumber \end{eqnarray}
   For $\Omega_c\to 1/R$ the minimal critical value, $V_{0c}\sim m^*/\sqrt{m^*R}\ll m^*$, is reached for $c_1\sim \sqrt{m^*R}\gg 1$.  For $(m^*R)^2\gg c_1\gg \sqrt{m^*R}$ we have
   \be
   V_{0c}\simeq 1.86 c_1^{1/3}m^*/(m^*R)^{2/3}\ll m^*\,.\label{Vc2}
    \ee
    Thus the minimal value of $V_{0c}$ is $V_{0c}\sim \sqrt{m^*/R}$ corresponding to $c_1\sim \sqrt{m^*R}$.
    The amplitude of the arising vortex field is limited by the redistribution of the charge, which we did not take into account assuming that $V_0\simeq const$. If we assumed conservation of the   angular momentum applied to the system rather than the angular velocity, the coefficient $c_1$ would be also limited by the conservation of the value of the angular momentum.

    If we used the relation $\phi^\prime (r=R)=0$, as the boundary condition, cf. Eq. (\ref{boundder}), we would arrive at the spectrum
\begin{eqnarray} \mu =-\Omega\nu -V_0
 +\sqrt{{m}^{*2}+
 j^{\prime\,2}_{1,\nu}/R^2}\,,
 \label{grlevprime}\end{eqnarray}
 as it is easily found with the help of the recurrence relations for the Bessel functions. For $\nu\gg 1$ condition (\ref{grlevprime})  differs from (\ref{grlev}) only in small  correction terms. Using Robin condition (\ref{boundderRob}) we see that at least for $c_1\gg 1$ the instability may occur  at
 \be
 1/R>\Omega >(1-\alpha)/R\,,\quad \alpha >0\label{Omder}
 \ee
  even for $V_0=0$.

In the comoving frame,   using Eq. (\ref{EnerPhipi}), 
we find that the  contribution of the individual vortex to the  ground state energy is
\begin{eqnarray}
&{\cal{E}}_{\pi} (\Omega)=\mu
 N_\pi
,
\label{bosenrot}
\end{eqnarray}
with $\mu$ given by Eq. (\ref{grlev}) for $n=1$ and
\begin{eqnarray}
&N_\pi=
2\pi d_z R^2 \sqrt{m^{*2}+\frac{j_{1,\nu}^2}{R^2} }\phi_{0}^2 J_{\nu +1}^2 (R/l_\phi)\,\label{NBes}
\end{eqnarray}
 having  sense of the  number of particles occupying the ground state level $\nu\neq 0$;  $\phi_0^2>0$ is arbitrary positive constant. To get Eq. (\ref{NBes}) we used the relation $\int_0^1 x dx J_\nu^2 (xR/l_\phi)=J_{\nu +1}^2 (R/l_\phi)
 /2$.
 The condensate of particles may appear from the vacuum provided the ground state level crosses zero, then we have
 \be{\cal{E}}_{\pi} (\Omega)<0\,.
  \ee
As we see from Eqs. (\ref{bosenrot}), (\ref{NBes}), for $\Omega>\Omega_c$  the amplitude $\phi_0$ of the field is not limited without taking into account of the self-interaction, $\lambda>0$, and/or redistribution of the charge and the magnetic field in the system.


Notice
that  individual vortices can form a lattice with distance between vortices $\delta R\gsim  l_\phi$, $R\gg \delta R$. Their total energy then is ${\cal{E}}_{\pi}^{\rm tot} (\Omega)\sim N_{\rm vort}{\cal{E}}_{\pi} (\Omega)$, where $1\ll N_{\rm vort}\lsim R^2/l_\phi^2$. The global charge neutrality condition is fulfilled provided the vortices (with winding number $\nu$, and charge $e$) and antivortices ($-\nu$, $-e$) alternate.
The pion field can be then presented, as sum of vortex solutions counted from  points $\vec{r}_n=(x_n, y_n)$, forming the lattice
$\phi (x,y)=\sum C_n \phi(x-x_n,y-y_n)$ with $C_n=C_{b+n}$; for the quadratic lattice $b=1$, for triangle lattice $b=2$.


As it was mentioned, we put $\vec{A}=0$   in Eq. (\ref{pipitot}). In this case the vortex current (\ref{current}) in the
comoving
 frame is nonzero and for $R>r\gg l_\phi$ it is given by
\be{j}_\theta^\pi\simeq 2e\nu |\phi|^2(1+O(e^2 l_\phi^2|\phi|^2))/r\,.\label{jlab}\ee
Some extra assumptions, which we have used to derive expressions presented in this section, are discussed in Appendix \ref{Assumptions}. 

\subsection{Case $H\to 0$  in presence of self-interaction }\label{HzeroSelfint}
The condensate energy has the minimum  for the static field, cf. Eq. (\ref{EnerPhipi}).
For $\lambda\neq 0$, $V_0\simeq const$, $H=0$ using the dimensionless variable $x=r/l_{\phi 0}$, where $l_{\phi 0}=1/\sqrt{\bar{\mu}_0^2-{m}^{*2}}$, i.e. $l_\phi$ at $\mu =0$, $\bar{\mu}_0= \bar{\mu}(\mu =0)= \Omega\nu + V_0$, employing Eq. (\ref{pipi})
 we arrive at equation
\be (\partial^2_x +x^{-1}\partial_x -\nu^2/x^2)\chi+\chi -\lambda\phi_{0}^2 l_{\phi0}^2\chi^3=0\,.\label{filvorteqdim11lambda}
\ee
 Choosing $\lambda\phi_{0}^2 l_{\phi0}^2=\lambda\phi_{0m}^2 l_{\phi0}^2=1$ we have
\begin{eqnarray}&\phi_{0m} =\sqrt{ (\bar{\mu}^2_0-{m}^{*2})/\lambda}\times\theta (\bar{\mu}^2_0-{m}^{*2}),\label{Phich2}
\end{eqnarray}
being the solution of Eq. (\ref{filvorteqdim11lambda}) at $x\to \infty$ satisfying the condition $\chi (x\to \infty)\to 1$. There exist  asymptotic solutions of Eq. (\ref{filvorteqdim11lambda}):  $\chi \propto x^{|\nu|}$ for $x\to 0$ and  $\chi =1-\nu^2/(2 x^2)$ for $x\gg \nu >0$.  Then  the field $\phi$ is expelled from the vortex core and the equilibrium value of the pion field (\ref{Phich2})
 is recovered at $r\gg \widetilde{l}_{\phi 0}=\nu l_{\phi 0}$. To fulfill the boundary condition $\chi (R/l_\phi)=0$ the solution should be modified only in a narrow region $R-r\sim l_{\phi 0}$ near the boundary. In case of the boundary condition $\chi^\prime_r (R/l_\phi)=0$, cf. (\ref{boundder}), such a modification is not required. Finally we note that in case $\mu=0$ the condition $R\gg \widetilde{l}_{\phi 0}$ is fulfilled  only for $c_1\ll 1/(l_{\phi 0}m^*)=\sqrt{(\Omega\nu+V_0)^2-m^{*2}}/m^*$.


In (\ref{filvorteqdim11lambda}), as in (\ref{pipi2}), we neglected the dependence on $\vec{A}$; for $\phi\sim \phi_{0m}$ the condition (\ref{estA}) of Appendix \ref{Assumptions}  holds for $\lambda\gg 8\pi e^2$. Evaluation of the magnetic field of the vortex line at 
$r\gg \widetilde{l}_{\phi 0}$ is performed in Appendices \ref{AppendixA} and \ref{AppendixB}. An alternative solution will be considered in Sect. \ref{Aharonov-sect}.

In the condensed matter physics the rotation velocities are tiny and energetically favorable are the vortices with $\nu =1$. In this case the length scales $l_{\phi 0}$ and $\widetilde{l}_{\phi 0}$ are of the same order of magnitude.  The important role is played by the Ginzburg-Landau parameter $\kappa=l_h/l_{\phi 0}$. For $\kappa<1/\sqrt{2}$ one deals with the superconductors of the second kind and for $\kappa >1/\sqrt{2}$, with the superconductors of the first kind. In case when the vortices are created  in the rapidly rotating empty vessel, at $\nu \gg 1$  we have $\widetilde{l}_{\phi 0}\gg  l_{\phi 0}$. Then  the
 role of the effective Ginzburg-Landau parameter is played by the  quantity $\widetilde{\kappa}=l_h/\widetilde{l}_{\phi 0}$, cf. Appendix \ref{AppendixA}.

 The pion condensate contribution to the energy for $\mu =0$ is as follows,
\begin{eqnarray} &{\cal{E}}_{\pi} (\Omega)=-2\pi d_z \int_0^R r dr \psi^*[\bar{\mu}_0^2+\hat{K} -m^{*\,2}]\psi\nonumber\\
 &+\lambda 2\pi d_z \int_0^R r dr|\psi|^4/2\,,
\label{EnGuo}\end{eqnarray}
with $\hat{K}$ given by Eq. (\ref{pipi}) taken at $H=0$, $\phi=e^{i\nu\theta-i\mu t}\psi$.
To evaluate the energy of the ground state $(1,\nu)$ level let us employ the variational procedure. For $R\gg \widetilde{l}_{\phi 0}$ that holds   for $c_1\ll \sqrt{\bar{\mu}_0^2-m^{*2}}/m^*$ the contribution of the region $r\lsim \widetilde{l}_{\phi 0}$ in the integral is narrow and for $r\gg \widetilde{l}_{\phi 0}$ we may put $\psi$ to be real constant. Variation of (\ref{EnGuo}) in $\psi$ yields $\psi =\phi_{0m}$, cf. (\ref{Phich2}). Thus we arrive at the energy
\begin{eqnarray}
&{\cal{E}}_{\pi} (\Omega)\simeq -\frac{\pi d_z R^2 [\bar{\mu}_0^2-m^{*2}]^2}{4\lambda}\theta(\bar{\mu}_0^2-m^{*2})\,.\label{probEnnoHrest1}
\end{eqnarray}
The critical angular velocity in this case  is given by the relation
\be \Omega_c^\pi =(m^*-V_0)/\nu\,.\label{omcpi}\ee

For  $R\lsim \widetilde{l}_{\phi 0}$ that holds   for $c_1\gsim \sqrt{\bar{\mu}_0^2-m^{*2}}/m^*$ the asymptotic solution $\chi\to 1$ is not reached and the typical distance, at which $\chi$ is changed, is $l_{\phi 0}$. In this case,  as a probe functions, it is convenient to use the eigenfunctions of the linear equation
\be
-\hat{K}\psi =K_{n,\nu} \psi\,\label{eigen}\ee
with $K_{n,\nu}=\bar{\mu}^2 -m^{*\,2}$, cf. Eqs. (\ref{pipi}), (\ref{filvorteqdim11lambda}) for $\lambda=0$. The eigenfunctions  are $\psi =C\chi_{\rm pr}(r)=CJ_\nu (r)\,,$
 $C$ is the real constant. As the boundary conditions, let us use (\ref{boundcond}) and (\ref{boundcond0}).
Then from (\ref{EnGuo})
after minimization in $C$ we find for $\bar{\mu}_0^2>m^{*2}+j_{1,\nu}^2/R^2$, cf. Eqs. (\ref{Vc1}), (\ref{Vc2}),
\begin{eqnarray}
&{\cal{E}}_{\pi} (\Omega)=-\frac{\pi d_z R^2 J_{\nu+1}^4(R/l_{\phi 0})[\bar{\mu}_0^2-m^{*2}-{j_{1,\nu}^2}/{R^2}]^2}{4I\lambda}\,,\label{probEnnoH}
\end{eqnarray}
where $I=\int_0^1xdxJ_\nu^4 (Rx/l_{\phi 0})$. The critical angular velocity is given by Eq. (\ref{critcentr}), which value is larger than that given by Eq. (\ref{omcpi}). Finally we notice that although Eq. (\ref{probEnnoH}) was derived for $R\lsim \widetilde{l}_{\phi 0}$, if we formally employed  it  for $R\gg \widetilde{l}_{\phi 0}$ and used  the asymptotic expression for the Bessel function, we would arrive, with a logarithmic accuracy, at an estimation $J_{\nu+1}^4(R/l_{\phi 0})/I\sim \frac{8}{3\ln (R/l_{\phi 0})}$. Although
in this case the energy gain given by Eq. (\ref{probEnnoH}) is $\sim 1/\ln (R/l_{\phi 0})\ll 1$ times smaller than that given by  (\ref{probEnnoHrest1}), it is interesting to notice that
the critical angular velocities coincide, being given by  Eq. (\ref{omcpi}).

\subsection{Case $N=eHR^2/2\geq \nu\geq 1$ in absence of self-interaction}\label{HNonint}
Let us continue to study the case $m^{*2}>0$. As for zero and a weak external magnetic field considered above, in case  of a strong external magnetic field studied in \cite{Zahed,Guo} the  current associated with the arising vortex proves to be  non-zero in the
local flat frame. For $\lambda =0$ and $V=-V_0=const$, $H\neq 0$,   the solution of the equation of motion (\ref{pipi}) renders, cf. \cite{LL3},
 \begin{eqnarray}
 \chi (r)=r^{|\nu|}{e^{-|eH|r^2/4}} _{1}F^1(-a, |\nu|+1, |eH|r^2/2)\,,\label{hypergeom}
\end{eqnarray}
 $_{1}F^1$ is a confluent (degenerate) hypergeometrical function and
 \be-a=\frac{1}{2}(|\nu|-\nu+1)-\frac{(\mu +\Omega\nu +V_0)^2-m^{*2}}{2eH}\,,\ee
 that for $\nu\geq 0$ can be rewritten as
  \begin{eqnarray}
 \mu =-V_0-\Omega\nu+\sqrt{m^{*2}+|eH|(1+2a(\nu))}\,.\label{dispH}
 \end{eqnarray}

The function $_{1}F^1$ can be expanded in the  series,
 \begin{eqnarray}
 _{1}F^1 (\alpha,\beta,z)=1+\frac{\alpha z}{\beta 1!}+\frac{\alpha (1+\alpha) z^2}{\beta (\beta +1)2!}+...\label{expF}
 \end{eqnarray}
 Finiteness of $ _{1}F^1 $ for all $r\in (0,\infty)$ requires $a_{n,\nu}=-\alpha$ to be integer nonnegative number, $n\geq 1$.
For $a =0$ one has $_{1}F^1=1$ and  the energy $\mu$ in (\ref{dispH}) gets the minimal value.
The condition $\mu < 0$ holds for
\be \Omega> \Omega_c^H=(-V_0+\sqrt{m^{*2}+|eH|})/\nu\,.\label{OmH}\ee
However the boundary condition (\ref{boundcond}) is not fulfilled in this case.

The quantity $N=|eH|S/(2\pi)=|eH|R^2/2$, where $S=\pi R^2$, has the meaning of the degeneracy factor of the level at given  $n,p_z$, $\nu\leq N$, cf. \cite{LL3}. For $N=|eH|R^2/2=\nu+1$, using the  boundary condition (\ref{boundcond}) we have $\alpha=-1$ in (\ref{expF}), that corresponds to $a=1$ in (\ref{dispH}), and  only first two terms survive in the series in this case. For $\alpha=-2$ three terms survive, etc.

Assuming that the point $r=R$, in which the boundary condition (\ref{boundcond}) is supposed to be fulfilled can be chosen arbitrary (certainly for  $r<1/\Omega$), Refs.  \cite{Zahed,Guo} allowed for noninteger values of $a_{n\nu}$. The noninteger minimal value of $a(\nu)$ decreases with increasing $N$ and for $R\to \infty$ one recovers  the  solution $a=0$.  However with noninteger values of $a$ the field   $|\phi|$ gets the maximum not at the physically meaningful value
  $r\sim  \sqrt{\nu/|eH|}\,,$
as it were for the integer nonnegative  $a$, but  in the vicinity of  the point $r=R$, in which  one artificially  imposed the boundary condition.
 It may have a physical sense only provided the value $R$ coincides  with the size of the system, e.g. the size of the  fireball in case of   heavy-ion collisions, or with the internal size  of the vessel in case of the rotating vessel, at the assumption that the systems rotate as the rigid body, cf. \cite{Voskresensky:2024ivv}. Then it looks natural to put the boundary condition not at arbitrary chosen $R$ but  at the radius  $R$ of the system  \cite{Voskresensky:2023znr,Voskresensky:2024ivv}. In case of the rotating vessel we assume that  the field does not penetrate the vessel's  walls. On the other hand, in
 case when the typical distance, at which the pion field is changed significantly is
\be R_{H,\nu}=\sqrt{\nu/|eH|}\ll 1/\Omega\,,\label{Larmor}\ee
 the field (\ref{hypergeom}) is already exponentially small at $r\sim 1/\Omega$ and at $r\sim R$ for $R\gg R_{H,\nu}$. Thereby one may hope that violation of the causality condition at such large distances will not influence  physical properties of the system. Then the ground state corresponds to $a=0$ in (\ref{dispH}).   Thus, if  the condition (\ref{Larmor}) is fulfilled, it seems  more physically motivated to put $a=0$ and do not care about boundary condition at $r=R\gg R_{H,\nu}$, cf.  \cite{Voskresensky:2023znr,Voskresensky:2024ivv}, rather than to employ the noninteger values of $a$. We see that  condition (\ref{Larmor}) is always fulfilled for $N\gg \nu$ for all  $\Omega<1/R$.

  For   $|eH|\ll m^{*2}$ from (\ref{OmH})  we obtain that $\Omega_c^H\simeq \Omega_c^\pi$ given by Eq. (\ref{omcpi}).
   The dependence on $R$  existed for $H=0$ in Eq. (\ref{grlev}) disappeared, cf. \cite{Voskresensky:2023znr,Voskresensky:2024ivv}. For $\Omega \nu\sim m^*$ of our interest the conditions $|eH|\ll m^{*2}$ and (\ref{Larmor}) are fulfilled for $(m^*R)^2\gg N\gg (m^*R)^2/\nu\gg 1$.

The pion energy is given by Eq. (\ref{bosenrot}), now with $\mu$ from (\ref{dispH}) and
\be
N_\pi=\mu d_z 4\pi\sqrt{m^{*2}+|eH|(1+2a)}\phi_{0}^2\int_0^R r dr  \chi^2\,,\ee
with $\chi$ from (\ref{hypergeom}).

\subsection{Case $N=eHR^2/2\gg 1$  in presence of self-interaction}\label{HSelfint}

Reference \cite{Guo} studied the case $N=eHR^2/2\gg 1$ numerically employing noninteger values of $a$.
Let us  consider analytically the case when the condition (\ref{Larmor}) is fulfilled and we may ignore presence of the boundary, as it has been discussed above.
The state of the minimal energy $\epsilon_{n,\nu}$   corresponds to $n=1$ and $a=0$ and    we have
$K_{1,\nu}=|eH|(|\nu|-\nu +1)/2$,
$\psi =C\chi_{\rm pr}(r) =C r^{|\nu|}e^{-|eH|r^2/4}$. This eigenfunction describes  an individual vortex characterized  by the winding number $\nu$.
From (\ref{EnGuo}) after minimization in $C$, using the Stirling formula $\nu!\simeq \sqrt{2\pi\nu}(\nu/e)^\nu$
we obtain expression
\be{\cal{E}}_{\pi} (\Omega)\simeq - \frac{2\pi^{3/2} d_z R^2_{H,\nu}[(\Omega\nu +V_0)^2-m^{*\,2}-|eH|]^2}{\nu^{1/2}\lambda}\ee
valid for $\Omega>\Omega_c^H$ and $N\gg \nu >0$. The larger is $\nu$ the smaller is $\Omega_c^H$.

The supervortices, being formed at distances $\delta r \gsim R_{H,\nu}$ from each other, can   form the lattice. Their number is $N_{\rm vort}\lsim R^2/R_{\rm H,\nu}^2$ and their energy is thereby estimated as ${\cal{E}}_{\pi}^{\rm tot} (\Omega)\sim N_{\rm vort}{\cal{E}}_{\pi} (\Omega)$.
The maximum value of the external magnetic field $H_{c2}$, below  which  there exists a  pion vortex condensate, is given by
\be |eH_{c2}| =1/l_{\phi 0}^2=(\Omega\nu +V_0)^2-m^{*\,2}>0\,.\label{hc2}\ee

\section{Aharonov-Bohm effect and absence of current in comoving frame}\label{Aharonov-sect}
In quantum mechanics owing to the Aharonov-Bohm effect  the charged particle can be affected by an electromagnetic potential, even being confined in the region, in which both the magnetic  and electric fields  are zero. Let us employ this fact in our problem. We take
 \be e\vec{A}_\theta =e\vec{A}^{\rm AB}_\theta +e\vec{A}^{\rm v}_\theta\,,\quad  e\vec{A}^{\rm AB}_\theta(r)=\nu/r\,.\label{Aharonov}
  \ee
  The magnetic field  corresponding to the auxiliary vector potential $\vec{A}^{\rm AB}_\theta$ in (\ref{Aharonov})  is $\vec{h}_z^{\rm AB}=(\mbox{curl} \vec{A}^{AB})_z=\frac{1}{r}\frac{\partial (rA_\theta)}{\partial r}=0$. The current density (\ref{current}) corresponding to fields (\ref{Phifieldformpi}) and (\ref{Aharonov}), as well as the combination $(\nabla_\theta -ie\vec{A}^{\rm AB}_\theta)\phi$, are also zero
  and the Maxwell equation (\ref{Meis}) is satisfied in case when $\vec{j}_{\rm ex}=0$ in the rotation frame. Especially one should point out that with such a field ansatz the centrifugal term $(\nu^2/r^2)\phi$ disappears from the equation of motion (\ref{pipitot})  and owing to this circumstance the asymptotic solution  $\chi \simeq 1$ is now reached already at the distance  $r\gg l_\phi$ from the vortex center rather than at $r\gg \widetilde{l}_{\phi 0}$.
Equation (\ref{Meis}) for $\vec{A}^{\rm v}_\theta (r)=a(r)/r$ then renders
\be a^{\prime\prime}-a^{\prime}/r -8\pi e^2 \phi_{0}^2\chi^2 a=0\,.\label{Eqa}\ee
The condition $\chi(r)\simeq const$ is also fulfilled at typical distances from  the boundary
  \be
  R-r\gsim l_h \sim l_\phi\kappa\gg l_\phi\,,\label{amag}\ee
where $\kappa=\sqrt{\lambda/(8\pi e^2)}$ is the  Ginzburg-Landau parameter.  At these distances in absence of the external magnetic field one may use the  trivial  solution of Eq. (\ref{Eqa}), $a=0$. For concreteness we shall assume that  $\kappa\gg 1$.

For $H\neq 0$  one should solve Eq. (\ref{Eqa}) employing the boundary conditions
\be a(0)=-\nu\,, \quad a^\prime (r=R)=H\,,\label{abound}
\ee
 for $r>R$ we use $\vec{A}=(-Hy/2, Hx/2, 0)$, also cf. Appendices \ref{AppendixA} and \ref{AppendixB} and \cite{Zharkov2001}.

For $H=0$, with the ansatz (\ref{Aharonov}) seeking solution in the form (\ref{Phifieldformpi}) with $\xi =\nu\theta$ and setting $a=0$, from Eq. (\ref{pipitot})  we arrive at equation
\be (\partial^2_x +x^{-1}\partial_x )\chi+\chi -\lambda\phi_{0}^2l_{\phi}^2\chi^3=0\,,\quad x=r/l_\phi\label{filvorteqdim11lambda1}
\ee
for the order parameter not containing the centrifugal term in the comoving frame, which existed in  Eq. (\ref{filvorteqdim11lambda}). Thus for $\lambda =0$ instead of Eq. (\ref{grlev}) we now obtain the spectrum
\begin{eqnarray} &\mu =-\Omega\nu -V_0
 +{m}^{*}\,,
 \label{grlevbohm}\end{eqnarray}
that results in the value of the critical angular velocity
\be\Omega_c^{\rm AB}=\Omega_c^\pi=\Omega_c^H(H=0)\,,\ee
  cf. Eqs. (\ref{omcpi}), and (\ref{OmH}) for $H=0$. Due to the compensation of the centrifugal term by the Aharonov-Bohm vector potential, the value $\Omega_c^{\rm AB}$  proves to be smaller than the value $\Omega_c$ given by Eqs. (\ref{critcentr}), (\ref{condVom}).

Now,  introducing the variable $u=(r-R)/l_\phi$ we have
\be \partial^2_u \chi+(u+R/l_\phi)^{-1}\partial_u \chi+\chi -\lambda\phi_{0m}^2l_{\phi}^2\chi^3=0\,.\label{filvorteqdim11lambdaAB}
\ee
For typical values $-u\sim 1$ we may drop the second term in (\ref{filvorteqdim11lambdaAB}). Then geometry becomes effectively flat and  the solution of Eq. (\ref{filvorteqdim11lambdaAB}) for $\mu=0$ satisfying the boundary conditions $\phi  (u=0)=0$, $\phi (-\infty)=\phi_{0m}$  yields
\be
\phi =-\phi_{0m} \mbox{th} (u/\sqrt{2})\,.
\ee
With this solution we arrive at  the energy given by Eq. (\ref{probEnnoHrest1}).

Comparing (\ref{probEnnoHrest1}) and (\ref{probEnnoH}) we see that it is energetically profitable for the vortex condensate to remain at rest in the
comoving frame rather than to form  a nonzero current there. In spite of this circumstance, the solution, which we have derived in Sect. \ref{HzeroNonint} and in \cite{Voskresensky:2023znr,Voskresensky:2024ivv}, is also  physically meaningful  due to the conservation of the winding number.  Which solution is realized,  depends on the dynamical conditions. Indeed, if a tiny vortex $\phi$ field was formed producing a weak current in the comoving frame, given by Eq. (\ref{jlab}), then it can  be  reconstructed to the more energetically favorable state of the developed condensate  corresponding to the zero current in the same frame  only owing to the presence of a  weak interaction with the walls (in case of the rotating vessel), which allows to absorb the surplus  winding number.

For $H\neq 0$, one has $a\neq 0$ and Eq.
(\ref{filvorteqdim11lambda1}) is replaced by
\be (\partial^2_x +x^{-1}\partial_x )\chi+\chi -\lambda\phi_{0}^2l_{\phi}^2\chi^3+e^2(A_\theta^{\rm v})^2l_{\phi}^2\chi=0\,, \label{filvorteqdim11lambda1H}
\ee
and $a$ satisfies Eq. (\ref{Eqa}) for $r<R$ and the boundary conditions (\ref{abound}).
Similar equations occurring in the theory of metalic superconductors  for $\Omega =0, H\neq 0$  were solved numerically, cf. \cite{Zharkov2001}. The case $\widetilde{\kappa}=l_h/\widetilde{l}_{\phi 0}\gg 1$ is considered in Appendix \ref{AppendixB}. The solution found in Sect. \ref{HNonint}  holds for $H$ in the vicinity of $H_{c2}$.

 \section{Meissner effect and  London law in relativistic region}\label{sect-London}

Now we come to the discussion of the London moment and Meissner effect.

\subsection{Rotation of  empty neutral vessel}

 If { rotated is  the empty neutral vessel}, $n_{\rm ex}=0$, $\vec{j}_{\rm ex}=0$, then  there is no reason for the appearance of  the London moment. If the external magnetic field is absent, for $\Omega >\Omega_c$ there may appear a supervortex with a current density in the comoving frame given by $j_\theta^\pi \simeq 2e\nu\phi_0^2/r$, cf. Eq. (\ref{jlab}). This case  was considered in Sects. \ref{HzeroNonint}, \ref{HzeroSelfint}. However already for  much lower values of the angular velocity, for $\Omega_c\gg\Omega >\Omega_c^{\rm AB}$, there may appear a supervortex resting in the comoving frame (with $j_\pi =0$). This case was analyzed in Sect. \ref{Aharonov-sect}.

 In Sects. \ref{HNonint}, (\ref{HSelfint}) we did not take into account possibility of
  the Meissner effect occurring at least in a sufficiently weak  external magnetic field. Approximation of the  uniform magnetic field   employed in Ref. \cite{Zahed}  at $\lambda\to 0$ proves to be  approximately fulfilled
only  for
$H\simeq H_{c2}(1-O(\lambda|\phi|^2 l_\phi^2))$, when  the condensate term  can be  considered as small in Eq. (\ref{pipi}), and provided
the condition $\Omega >\Omega_c^H (H=H_{c2})$ is fulfilled, cf. Eq. (\ref{OmH}). Moreover, this solution holds only for $N(H_{c2})\sim |eH_{c2}|R^2\gg 1$ and  for $\kappa \gg 1$.
In the latter case there exists a region of  fields, $H_{c1}<H<H_{c2}$, in which   we deal either with the supervortex or with the lattice of  vortices, see derivations in Appendices \ref{AppendixA} and  \ref{AppendixB}.  For $H<H_{c1}$ (for $\Omega >\Omega_c^H (H=0)$) the magnetic field is fully expelled from the vessel due to the Meissner effect.
\\
 \subsection{Charged boson condensation in rapidly rotating piece of nuclear matter}

Another important case is  the rotation of the piece of the  matter consisting charged components. In  case of ordinary superconductors one deals with the ion-electron electrically neutral lattice. In  case of  the  nuclear  matter, when the $\pi^-$ level reaches zero,   the $\pi^-$ condensate can be formed, cf. \cite{Voskresensky:2023znr}. Again,  such a system is electrically neutral except a surface layer.
For irrotational motion  ($\nu =0$) the superfluid pion current in the laboratory frame is given by
$\vec{j}_\pi^{\rm lab} =-2e^2\vec{A}^{\rm lab}|\phi|^2_{\rm lab}$, cf. Eq. (\ref{current}) for $V_0^2-m^{*2}>0$.
The ground state  with $\phi^{\rm lab}\neq 0$ corresponds to $\nu=0$ in this case.

For $\nu\neq 0$, setting ${e}\vec{A}^{\rm lab}=\nu/r +{e}\vec{A}_{\rm L}^{\rm lab}$   in Eq. (\ref{current}), in the laboratory frame for $r\gg \nu l_\phi$   we obtain  $\vec{j}^{\rm lab}_\pi =-2e^2\vec{A}_{\rm L}^{\rm lab}|\phi_0^{\rm lab}|^2$.
Similarly to Eq. (\ref{curMacro}), in case of the   nuclear system    for $r\gg \nu l_\phi^{\rm lab}$ we put $\vec{j}^{\rm lab}=\vec{j}^{\rm lab}_\pi +\vec{j}^{\rm lab}_p=0$, where (in the cylindric geometry as a simplification) we have $\vec{j}_p^{\rm lab}=-en_p[\vec{\Omega}\times\vec{r}_3]$, and using that $n_p = n_\pi=2\bar{\mu}^{\rm lab}|\phi^{\rm lab}_0|^2=2\bar{\mu}|\phi_0|^2$ we arrive at the expressions describing the London moment in the relativistic case,
\be
e\vec{A}_{\rm L}^{\rm lab}=-\bar{\mu}^{\rm lab}[\vec{\Omega}\times\vec{r}_3]\,,\quad \vec{h}^{\rm lab}_{\rm L}=-(2\bar{\mu}^{\rm lab}/e)\vec{\Omega}\,,
\ee
$\bar{\mu}^{\rm lab}=\mu+V_0$, $V_0$ is the solution of the charge neutrality condition. In the nonrelativistic limit $\mu\simeq m^*$, $\bar{\mu}\simeq m^*$, and we recover the known result (\ref{London}).
In case of the $\pi^-$ condensate formed in the piece of the rotating nuclear matter we should put $\mu = 0$.
In the
comoving frame the normal current is absent and the London field is $\vec{A}_{\rm L}=0$.

As it was argued in Ref. \cite{Voskresensky:2023znr}, in case of  a rotating     nuclearite at the condition  $1/R>\Omega\gsim m/\nu$, i.e. for $1>\Omega R>1/c_1$,
it might be profitable to form a charged pion   vortex condensate, which will stabilize the system in the
comoving  frame. The kinetic energy of such a rotating nuclear systems is then lost  via a surface electromagnetic radiation. In case of a very large number of baryons such a radiation is strongly suppressed and  the rapidly rotating system is long-living.

\section{Conclusion}
This paper studied occurrence of the instability in the  charged pion vacuum resulting in the pion condensation  within the $\lambda|\phi|^4$ model at the rapid rotation of the system
placed in the scalar potential well,  under action of the electric and magnetic fields.
 The problem of the instability of the charged pion field  in the rotation frame was recently 
 studied in Refs. \cite{Zahed,Guo,Voskresensky:2023znr,Voskresensky:2024ivv}. References \cite{Zahed,Guo} studied behavior of the charged pion vacuum   in the rotation frame placed in the strong uniform constant
magnetic field. References \cite{Voskresensky:2023znr,Voskresensky:2024ivv}  focused attention on the charged pion condensation in rapidly rotating systems including  effects of the electric potential well. The subtlety  of the problem is that the  production of the condensate field and particles depends on the noninertial reference frame, in which the phenomenon is described. In this sense the phenomenon is similar to the Unruh effect resulting in production of soft photons in the unevenly moving frame.
In the given work we included possibility of presence (or absence) of the external uniform magnetic field and square electric and scalar potential wells and possibility of redistribution of internal electric and magnetic fields and their influence on the vortex charged pion field arising from the vacuum in the rotating systems. The key point of the present work is that  in presence of the rotation and the uniform external magnetic field the charged pion vacuum and a piece of nuclear matter  behave similarly to a superconductor.  Thereby one should additionally take into account the Meissner effect and the London moment, which were not included in mentioned above works, whereas these effects play central  role in case of ordinary superconductors.

In Sect. \ref{chir} we introduced the Lagrangian of the model in the rotation frame and equations of motion for the charged pion field and static electric and magnetic fields.
 It was noticed that the equation for the vector potential of the magnetic field in the comoving
 frame acquires an anti-screening and magnetic susceptibility contributions to the current density proportional to $\Omega^2$.
Then we  discussed the role of  the choice of the boundary conditions at some point $r=R$ and the causality condition $R<1/\Omega$.  It would be rather doubtful, if the physically meaningful result depended on an arbitrary chosen value $R$. Thereby, to avoid this subtlety  following Refs.  \cite{Voskresensky:2023znr,Voskresensky:2024ivv} we associated the rotation frame in this case with the rigidly rotating vessel with $R<1/\Omega$, being  equal to  the internal radius of the vessel. Then, for instance, usage of the Dirichlet boundary condition  at $r=R$ means that the produced pion field does not penetrate through the wall of the vessel.

In Sect. \ref{sect-ignor}  the question was  studied about instability of the charged pion vacuum  in the
comoving frame  at ignorance of the Meissner effect and the London moment.    We considered the possibility of the creation from the vacuum  of the charged pion field supervortex in the local flat frame characterized by a large winding number $\nu>0$.
First, we suggested that there is no external uniform magnetic field $H$ or, if it is, the quantities $|eH\nu|, (eH)^2R^2$ are much smaller than either $\nu^2/R^2$ or  $|m^{*2}|$, and the internal magnetic field $h$ and vector potential can be neglected. The latter statement holds provided the vortex condensate field remains  rather weak. With the pion vortex there is associated a weak current density in the local flat frame for $|\phi|^2\ll 1/l_\phi^2$, cf. Eqs. (\ref{jlab}), (\ref{estA}), $l_\phi$ is the coherence length given in Eq. (\ref{Bessel}). A very weak own magnetic field of the vortex decreases exponentially at $r\ \gg l_h$, as it was demonstrated in Appendix \ref{AppendixA}, $l_h$ is the magnetic field penetration depth given in Eq. (\ref{lhlength}).
We briefly  repeated the key results of \cite{Voskresensky:2023znr,Voskresensky:2024ivv} and additionally considered  the case $m^{*2}<0$.

For  $m^{*2}<0$ the vacuum is unstable for creation of the charged pion field already in absence of the external electric potential, for $V=0$, and for $\Omega=0$ (for $\nu=0$). Stability is provided by the self-interaction, $\lambda>0$.

For $m^{*2}>0$, ignoring for simplicity possible redistribution of the electric field in presence of the vortex field, employing the Dirichlet boundary condition (\ref{boundcond}) we estimated the   minimal value of the electric potential $V_{0c}$ necessary for the appearance  of the supervortex, $V_{0c}\sim m^*/\sqrt{m^*R}\ll m^*$. The corresponding  optimal winding number of the supervortex proved to be   $\nu \sim ({m^*R})^{3/2}\gg m^*$. Additionally we studied  influence of the choice of the boundary conditions on the resulting pion spectrum for $\lambda =0$ (compare results (\ref{grlev}), (\ref{grlevprime}) derived employing the boundary conditions (\ref{boundcond}) and (\ref{boundder}), respectively). In case of the Robin boundary condition (\ref{boundderRob}), which is often employed, e.g., in the theory of superconductivity, for a very rapid rotation the instability may arise even for the external electric potential  $V_0=0$, cf. Eq. (\ref{Omder}).

Then for $m^{*2}>0$, employing the variational method   we evaluated   the condensate field energy in the overcritical region for $\lambda\neq 0$. For that we considered two cases, $R\gg \widetilde{l}_{\phi }=\nu l_\phi$ and $R\lsim \widetilde{l}_{\phi }$. In the former case the condensate field reaches constant value for $r\gg \widetilde{l}_{\phi }$, as it occurs for ordinary superconductors of the second kind, but now with the Ginzburg-Landau parameter
 $\widetilde{\kappa}=l_h/\widetilde{l}_{\phi } \gg 1$. The critical rotation angular velocity, at which there arises the pion supervortex, is given by Eq. (\ref{omcpi}) and the energy gain is given by  Eq. (\ref{probEnnoHrest1}).   For $R\lsim \widetilde{l}_{\phi 0}$ the critical rotation angular velocity is given by Eq. (\ref{critcentr}). In this case we deal with the superconductivity of the second kind  for the Ginzburg-Landau parameter ${\kappa}=l_h/l_\phi \gg 1$. However the condensate field decreases at $r\gg l_\phi$ rather than reaches constant value.
Employing, as the trial  function,  the solution   (\ref{Bessel}) of the linear equation of motion (\ref{pipi2}), we estimated the energy gain due to  appearance   of the pion supervortex, cf. Eq.  (\ref{probEnnoH}).

Then, in case $N=eHR^2/2\gg 1$ we  briefly  repeated the key results of \cite{Zahed,Guo} and additionally discussed  subtlety associated with the choice of the boundary condition  (\ref{boundcond}) at arbitrary $R<1/\Omega$. Arguments were given in favor of that   one may not care about the boundary condition at $r=R$  provided the inequality  (\ref{Larmor}) is fulfilled and, if inequality (\ref{Larmor}) is not satisfied,  one should impose the boundary condition at the internal radius  of the vessel. Additionally, employing the variational method   we evaluated   the  condensate field energy in the overcritical region for $\lambda\neq 0$, cf. Eq. (\ref{probEnnoH}). As the trial function, we used the solution  (\ref{hypergeom}) of the linear equation of motion. Arguments were given in favor of the  appearance of the lattice of supervortices in the given case, that differs from the case of nonrelativistic superfluids, where the choice of the winding number $\nu=1$ is energetically favorable.

Section \ref{Aharonov-sect} employs the fact that owing to the Aharonov-Bohm effect  the charged particle can be affected by an electromagnetic potential, even being confined in the region, in which both the magnetic  and electric fields  are zero. We selected the Aharonov-Bohm vector potential, as $e\vec{A}_\theta=\nu/r$, that corresponds to zero contribution to the magnetic field but allows to compensate the centrifugal term in the equation of motion for the vortex condensate. With such a solution in absence of the external magnetic field at distances from the vortex center $r\gg l_\phi$ the  current density vanishes in the
comoving frame.
We demonstrated that it is energetically profitable for the vortex condensate to remain at rest in the
comoving frame rather than to form  a nonzero current there.

Then in Sect. \ref{sect-London} we first considered  the case of the rotated  empty neutral vessel, at conditions $n_{\rm ex}=0$, $\vec{j}_{\rm ex}=0$. In this case  there is no reason for the appearance of  the London moment. In presence of the external uniform magnetic field for $H<H_{c1}$ (for $\Omega >\Omega_c^H (H=0)$) the magnetic field proves to be fully expelled from the vessel due to the Meissner effect. We argued that the
approximation of the  uniform magnetic field  employed in Ref. \cite{Zahed} for the description of the charged pion field in the vacuum in the local flat frame  proves to be  approximately satisfied
only  for $H\simeq H_{c2}(1-O(\lambda|\phi|^2)l_\phi^2)$, when  the condensate term  can be  considered as tiny, at
the condition $\Omega >\Omega_c^H (H=H_{c2})$, cf. Eq. (\ref{OmH}). Moreover this solution holds only for $N(H_{c2})\sim |eH_{c2}|R^2\gg 1$ and  for $\kappa \gg 1$.
For the   fields varying within the interval $H_{c1}<H<H_{c2}$ we deal either with  the supervortex or with the lattice of the supervortices.

Then we considered   the rapidly rotating piece of the nuclear matter. It was demonstrated that it is  energetically preferable to have zero net current in the
co-moving frame that results in  the solution $\vec{A}_{\rm L}=0$ at $\vec{A}_{\rm AB}\neq 0$. In the laboratory frame it leads to the appearance of the own uniform magnetic field $\vec{h}^{\rm lab}_{\rm L}=-(2\bar{\mu}^{\rm lab}/e)\vec{\Omega}\,$, associated with the London moment in the relativistic case.

Appendix \ref{sect-rotSKG} discusses subtlety of the transition from the  Klein-Gordon-Fock equation in the rotation frame to the Schr\"odinger equation in the nonrelativistic limit.
Also we introduced
the Schr\"odinger equation in the rotation frame  with the help of the local Galilei transformation. It was demonstrated that the terms  $\propto [\vec{\Omega}\times\vec{A}]$
do not arise in the former method, whereas they appear in the latter one. Thereby we may conclude that such a terms  are beyond the applicability of the nonrelativistic approximation and should be disregarded in case when one uses the local Galilei transformation in the phenomenological theory of rotating and unevenly moving superconductors.
However we stress that in  case of the neutral complex order parameter in the nonrelativistic limit both methods yield the same results.

Some details of calculations done in the paper body were deferred to Appendices  \ref{Assumptions}, \ref{AppendixA} and \ref{AppendixB}.

Above we studied  instabilities under the rotation for the charged pion vortex field (for $\nu\neq 0$). For the $\pi^0$, the instability in respect to the formation of the chiral wave under the rotation at $\nu\neq 0$ was  discussed in \cite{Voskresensky:2023znr} within the $\sigma$ model. Focusing on the study of the vacuum we fully ignored a possibility of an additional interaction  via the Wess-Zumino-Witten axial anomaly term,  cf. \cite{SonStephanov,Yamamoto,TeryaevZakharov,Eto:2023wul} and refs. therein.  For example,  the  inhomogeneous $\nabla\pi^0$ field may interact with $\Omega$ via the term $\propto \mu_N\mu_I \vec{\Omega}\nabla \pi^0$, where $\mu_N$ and $\mu_I$ are  the baryon and isospin chemical potentials, cf. \cite{Yamamoto}. Such effects may contribute   in case of a rotating piece of the nuclear matter already for $\nu =0$. These interesting issues will be considered elsewhere.

Concluding,  instability of the  charged pion vacuum resulting in the pion condensation was studied within the $\lambda|\phi|^4$ model at the rapid rotation of the system placed in the scalar potential well   under action of the electric and magnetic fields. The  Meissner and Aharonov-Bohm effects and the London moment were taken into account.

\acknowledgments
I thank E. E. Kolomeitsev  for fruitful discussions.
\\

\appendix
\section{Schr\"odinger and Klein-Gordon-Fock equations in rotation frame}\label{sect-rotSKG}
In the nonrelativistic case the transition from the laboratory frame to the rotation frame  in the Schr\"odinger equation is usually performed with the help of the local Galilei transformation in the rhs, cf. \cite{Fischer:2003zz},
\be \frac{\Delta }{2{m}^*}\Psi \to \left[\frac{(\nabla -i {m}^*\vec{W})^2}{2{m}^*}{\rev{+}}\frac{{m}^*\vec{W}^2}{2}\right]\Psi\,,
\label{rotSchrod}\ee
where $\vec{W}=[\vec{\Omega}\times \vec{r}_3]$.
The l.h.s of the Schr\"odinger equation, $(i\partial_t-V)\Psi$, where  $V=qA_0$, does not change. On the other hand in the general relativistic case the rotation is introduced employing tetrad (\ref{tetrades})
that results in the replacement $\partial_t\phi\to (\partial_t +y\Omega \partial_x -x\Omega \partial_y)\phi$ in the Klein-Gordon-Fock equation, cf. Eq. (\ref{shiftrotelectric}), whereas the spatial derivative term $\partial_i^2\phi
$ does not change. Thus there appears question can one  in nonrelativistic limit   reproduce the result (\ref{rotSchrod}) starting from the Klein-Gordon-Fock equation.

Let us for simplicity focus attention on the stationary case. Then, taking  $\phi=e^{-i\mu t}\phi_{\rm st}$  and  performing the replacement
$\mu\to m^*+E_{n.r.}$ in Eq. (\ref{murep}) and dropping small quadratic terms  $\sim E_{n.r.}^2$, $\Omega^2$, $V^2$ and corresponding cross-terms, employing Eq. (\ref{murep}) we obtain relation
\begin{eqnarray}
(\widehat{\widetilde{\mu}}^2-m^{*2})\phi\simeq 2m^* (E_{n.r.} -V +i\vec{\Omega}
[\vec{r}_3\times \nabla])\phi_{\rm st}\,,\end{eqnarray}
which  after usage that $\vec{\Omega}[\vec{r}_3\times \nabla]=\vec{W}\nabla$, $\nabla \vec{W}=0$ leads us to Eq. (\ref{rotSchrod}),  cf.  \cite{Voskresensky:2023znr}.

 In presence of the  magnetic field  one should additionally perform the ordinary gauge replacement $\nabla\to \nabla -iq\vec{A}$ in (\ref{rotSchrod}) and in the Klein-Gordon-Fock equation. We see that the terms $\propto [\vec{\Omega}\times \vec{A}]$ do not appear in the Klein-Gordon-Fock Eq. (\ref{pipitot}), whereas such contribution  arises after performing  the gauge replacement  in   Eq. (\ref{rotSchrod}), cf. e.g. \cite{Babaev:2013tha,Wang2020}. Thus we conclude that the cross-term $\propto [\vec{\Omega}\times \vec{A}]$ emerging in the expression $(\nabla -iq\vec{A}{\rev{-i}}m[\vec{\Omega}\times \vec{r}_3])^2$ is beyond the applicability of the nonrelativistic approximation.  Implying  existence of this  contribution, Ref. \cite{Babaev:2013tha} derived relations between critical values of the magnetic field $H_{c1}$, $H_{c2}$ and the critical rotation velocities $\Omega_{c1}$ and $\Omega_{c2}$. As we see, such analogy is not traced from the Klein-Gordon-Fock equation presented in terms of the quantities measured in the rotation frame. Thus  employment of the gauge transformation together with the local Galilei  transformation may lead to  unjustified results. However for the neutral complex scalar field in the nonrelativistic limit both methods yield the same results and one can safely use the local Galilei transformation.
\\
\section{Some assumptions used  in Sect. \ref{HzeroNonint}}\label{Assumptions}
Here we discuss some assumptions, which we have used to derive expressions of Sect. \ref{HzeroNonint}.

$\bullet$
 As it follows from Eq. (\ref{Meis}) for $j_{\rm ext}=0$, we have $|e\vec{A}_\theta (r)|\sim 8\pi e^2|\phi (r)|^2\nu l_\phi$ for $r\gsim l_\phi$ and thereby $A$-dependence  can be  neglected in Eq. (\ref{pipitot}) at least for
 \be|\phi_0|^2\ll 1/(l_\phi^2 8\pi e^2)\,.\label{estA}\ee
In Sect. \ref{HSelfint} we show  that $1/l_\phi^2=H_{c2}$, where $H_{c2}$ has the sense of the upper critical magnetic field in case when the system is placed in the external uniform magnetic field, cf. Eq. (\ref{hc2}). The terms, which we disregarded putting $\vec{A}=0$ in Eq. (\ref{pipitot}),   yield the contribution   to the energy $\propto |\phi|^4$, whereas the contribution (\ref{bosenrot}), (\ref{NBes}) is dominant for small $\phi$, being  $\propto \phi^2$. Then, provided inequality (\ref{estA}) is fulfilled, one indeed may put $A_\theta \simeq 0$,
  as it has been assumed in \cite{Voskresensky:2023znr,Voskresensky:2024ivv}.

$\bullet$ We solved Eq. (\ref{pipitot}) in assumption that $V$ is the external potential of the form $V\simeq -V_0=const$ for $r<R$,  whereas the charge can be  redistributed following Eq. (\ref{densmod1rot}). In absence of the external potential we may consider two cases: when  $n_{\rm ex}=const$ and $n_{\rm ex}=0$, cf. \cite{Voskresensky:2023znr}. In case of the rotating  system with $n_{\rm ex}=const$ for $l_\phi\ll r<R$, the solution of Eq. (\ref{densmod1rot}) is indeed $V\simeq const$ for $R-r\gg l_\phi$, corresponding to the local charge neutrality. In case of the rotating empty neutral vessel, inside the vessel the electric field is redistributed between the field of the supervortex and  the anti-vortex, the latter is pushed closer to the surface of the vessel. As it follows from the Poisson equation, the change of the potential
on the typical length $l_\phi$ is as small as $\delta V/V\sim l_\phi/R$.

$\bullet$ One may use the linear Eq. (\ref{pipi2}) also for $\lambda\neq 0$  provided $\lambda|\phi_0|^2 l_\phi^2\ll 1$,  cf. Eq. (\ref{pipitot}) and Eq. (\ref{filvorteqdim11lambda}) in Sect. \ref{HzeroSelfint}.
 Another, more energetically profitable  solution corresponding to the absence of the total current in the rotation frame, is considered in Sect. \ref{sect-London}.

$\bullet$ In case when the external magnetic field $H\neq 0$ we may neglect the $H$ dependence in the equation of motion (\ref{pipi}) provided  smallness of quantities $|eH\nu|, (eH)^2R^2$ compared with either $\nu^2/R^2$ or  $m^{*2}$.

$\bullet$  In assumption of the absence of the radiation and/or  absorption of antiparticles produced together with particles with the energy $\epsilon_{1,\nu}\simeq 0$ in reactions on the walls of the vessel, the instability occurs  when the ground state energy of the particle, $\epsilon_{1,\nu}$, reaches the boundary of the lower continuum  $-m$, cf. \cite{Zahed}. In case, if we deal with the gas of particles with the   fixed  (or dynamically fixed) particle number, the value  $\epsilon_{1,\nu}=\mu>0$ is determined by the particle number conservation at the time interval under consideration. In the latter case  ${\cal{E}}_{\pi} (\Omega)>0$.
\\

\section{Magnetic field and energy of the vortex line}\label{AppendixA}
Here we consider the charged pion vacuum in the co-moving local flat frame in absence of the external magnetic field $H\neq 0$, however we take into account the own magnetic field of the pion vortex $h\neq 0$.  In case   $R\gsim l_h\gg \widetilde{l}_{\phi 0}=\nu l_{\phi 0}$  for typical distances from the center of the vortex line  $r\sim l_h\gg \widetilde{l}_{\phi 0}$ we may use asymptotic solution of Eq. (\ref{filvorteqdim11lambda}) with  $\chi\simeq 1$ corresponding to the developed condensate. The current density is nonzero for $r\lsim l_h$ and drops exponentially for $r\gg l_h$, as we shall see. In case when the condensate field is rather weak, cf. condition (\ref{estA}) and condition $\phi\ll \phi_{0m}$,  it is sufficient to  employ only that $R\gsim l_h\gg l_{\phi 0}$.

In general, the further derivation given below is similar to that performed  for nonrelativistic superconducting materials  of the second kind for $\nu\sim 1$, cf.  \cite{LP1981}. This solution corresponds to assumption that $\vec{A}_{\rm AB}=0$ in Eq. (\ref{Aharonov}) of Sect. \ref{Aharonov-sect}.
Employing the Maxwell equation $\vec{j}_\pi =\mbox{curl}\vec{h}/(4\pi)$ and Eq. (\ref{current}),  for $r\gg \widetilde{l}_{\phi 0}$  we find
\be \int \mbox{curl}\vec{h}d\vec{l}=16\pi^2 e \nu \phi_0^2-\int \vec{h}d\vec{s}/l_h^2\,,\label{curlcurl}\ee
with the penetration depth of the magnetic field determined as
\be l_h^2=1/(8\pi e^2\phi_{0m}^2)\,.\label{lhlength}\ee
Integrations are done over the closed circular contour taken at $r\gg \widetilde{l}_{\phi 0}$, and the surface covering the contour.
Employing the Stokes theorem from (\ref{curlcurl}) we find
\be \int (\vec{h}+l_h^2 \mbox{curl}\mbox{curl}\vec{h})d\vec{s}=2\pi\nu/e\,.\label{hs}\ee
Since the circular contour taken at $r\gg \widetilde{l}_{\phi 0}$  is arbitrary, we arrive at the relation
\be \vec{h}+l_h^2 \mbox{curl}\mbox{curl}\,\vec{h}=\vec{h}-l_h^2 \Delta \vec{h}=2\pi\nu\delta(\vec{r})/e\,,\label{hrotrot}\ee
where $\int\delta(\vec{r})ds=1$, $\vec{l}[\nabla\times \vec{h}]=-\partial h_z/\partial r$.

For distances  $l_h\gg r\gg \widetilde{l}_{\phi 0}$  we may neglect the term
$\int \vec{h}d\vec{s}$ \,in (\ref{hs}) containing only a small part of the magnetic flux $\propto  r^2/l_h^2$, what is equivalent to condition of the dropping of the term $2e^2A_\theta |\phi|^2$ in the current density (\ref{current}) employed in Sect. \ref{HzeroNonint} in the derivation of the solution (\ref{Bessel}). Then  we obtain
\be -\partial h_z/\partial r \simeq\nu/(rl_h^2 e)\,,\ee
and thereby
\be e h(l_h\gg r\gg \widetilde{l}_{\phi 0})=\nu\ln(l_h/r)/l_h^2 \,\label{Mac}.\ee

Including the term $\propto h$, from (\ref{hrotrot}) we recover
\be h_z(r)=C_1 K_0 (r/l_h)\,,\label{Macdonald}\ee
 where $C_1$ is arbitrary constant and $K_0$ is the Macdonald function, $K_0(x\ll 1)\simeq -\ln (x\gamma/2)$, $\gamma =e^C$, $C=0.577...$, $K_0(x\gg 1)\simeq \sqrt{\pi/(2x)}e^{-x}$.
Matching (\ref{Macdonald}) and (\ref{Mac}) yields
\be
eh_z(r\gg \widetilde{l}_{\phi 0})=\nu K_0 (r/l_h)/l_h^2\,.
\ee
The vortex contribution to the  energy is as follows
\begin{eqnarray}
&{\cal{E}}_{\rm kin}^{(1)}\simeq  {\int d^3 X |(\nabla +ie\vec{A})\phi|^2} \simeq l_h^2\int
(\mbox{curl}\vec{h})^2 d^3 r_3 \nonumber\\
&=2\pi\nu^2 \phi_0^2 d_z \ln (l_h/\widetilde{l}_{\phi 0})\,,\label{Evortex11}
\end{eqnarray}
that within the logarithmic accuracy coincides with the solution  (30) in \cite{Voskresensky:2024ivv}, where dependence on $\vec{A}$ in this expression was neglected.
There is still the term ${\cal{E}}_h=\int \vec{h}^2 d^3 r_3/(8\pi)$ in the energy depending on $\vec{A}$, which  however is $1/\ln (l_h/\widetilde{l}_{\phi 0})\ll 1$ times smaller than the term (\ref{Evortex11}). With these approximations with the logarithmic accuracy  we recover results derived in \cite{Voskresensky:2023znr,Voskresensky:2024ivv} and used in Sects. \ref{HzeroNonint}, \ref{HzeroSelfint}.

Note that for $H=0$ the alternative solution  $\vec{A}_{\rm AB}=\nu/r$, cf. Eq. (\ref{Aharonov}) of Sect. \ref{Aharonov-sect}, and  $a=0$ satisfying Eq. (\ref{Eqa}), corresponding to $\vec{j}=0$ for $r\gg {l}_{\phi 0}$,  is more energetically favorable than the solution found  in Sect. \ref{sect-ignor} and in this Appendix.  However in presence of the external magnetic field situation changes as it will be argued in Appendix \ref{AppendixB}.
\\

\section{Mixed state. Critical values of magnetic field}\label{AppendixB}

Now we include external uniform magnetic field $\vec{H}$. From the expression for the Gibbs free energy,
\be G=E-\vec{M}\vec{H}\,,\quad 4\pi \vec{M}=\vec{h}-\vec{H},
\ee
 employing that  $\int\vec{h}d\vec{s} =\nu 2\pi/e$ and Eq. (\ref{Evortex11}) we are able to find the lower critical value of the external magnetic field, above which there arise  first vortices
 \be eH_{c1}=(\nu \ln\kappa)/l_h^2=8\pi e^2 \nu  \phi_{0m}^2\ln\widetilde{\kappa}\,\ee
for $(\Omega\nu+V_0)^2-m^{*2}>0$, $\widetilde{\kappa}=l_h/\widetilde{l}_{\phi 0}$.

For $H>H_{c1}$ with increasing $H$ there appears the lattice of vortices and the distance between them decreases. The  lattice disappears for $H>H_{c2}$. For $H$ in the vicinity of $H_{c2}$ the field $\phi$ becomes tiny and the term $\phi|\phi|^2$ in the equation of motion can be dropped, whereas the external magnetic field can be considered as approximately uniform. In this case considerations performed in  \cite{Zahed} and in Sect. \ref{HNonint} are valid. If  rotated is   empty neutral vessel, in case when $n_{\rm ex}=0$, $\vec{j}_{\rm ex}=0$, then  there is no reason for the appearance of  the London moment. If the external magnetic field is absent, for $\Omega >\Omega_c$ there may appear a supervortex with a current density in the co-moving frame given by $j_\theta^\pi \simeq 2e\nu\phi_0^2/r$, see Eq. (\ref{jlab}). However already for a much lower values of the angular velocity,  $\Omega >\Omega_c^{\rm AB}$, there may appear a supervortex resting in the
co-moving frame (with $j_\pi =0$), cf. Sect. \ref{Aharonov-sect}.
\\


\end{document}